\documentclass[oldversion]{aa}
\usepackage{epsfig}
\usepackage[switch]{lineno}
\usepackage{graphicx}
\usepackage{txfonts}
\usepackage{natbib}
\usepackage{url}

\bibpunct{(}{)}{;}{a}{}{,} % to follow the A&A style

\begin{document}

\title{The cooling rate of neutron stars after thermonuclear shell flashes}
\authorrunning{in 't Zand, Cumming, Triemstra, Mateijsen \& Bagnoli}

\author{J.J.M.~in~'t~Zand\inst{1}, A. Cumming\inst{2},
  T.L. Triemstra\inst{1}, R.A.D.A. Mateijsen\inst{1,3} \&
  T. Bagnoli\inst{1,4}}

%\offprints{J.J.M. in 't Zand, email {\tt jeanz@sron.nl}}

\institute{     SRON Netherlands Institute for Space Research, Sorbonnelaan 2,
                3584 CA Utrecht, the Netherlands; {\tt jeanz@sron.nl}
            \and
                Physics Dept., McGill University, 3600 Rue University,
                Montreal, QC, H3A 2T8, Canada
           \and
                Reynaertcollege, Postbus 32, 4560 AA Hulst, Zeeland, the
                Netherlands
           \and
                Astronomical Institute 'Anton Pannekoek', University of
                Amsterdam, Science Park 904, 1098 XH Amsterdam, The Netherlands
          }

\date{\it Recommended for publication, dd. Dec. 15th, 2013}

\abstract{Thermonuclear shell flashes on neutron stars are detected as
  bright X-ray bursts. Traditionally, their decay is modeled with an
  exponential function. However, this is not what theory predicts. The
  expected functional form for luminosities below the Eddington limit,
  at times when there is no significant nuclear burning, is a power
  law. We tested the exponential and power-law functional forms
  against the best data available: bursts measured with the
  high-throughput Proportional Counter Array (PCA) on board the Rossi
  X-ray Timing Explorer. We selected a sample of 35 'clean' and
  ordinary (i.e., shorter than a few minutes) bursts from 14 different
  neutron stars that 1) show a large dynamic range in luminosity, 2)
  are the least affected by disturbances by the accretion disk and 3)
  lack prolonged nuclear burning through the rp-process. We find
  indeed that for every burst a power law is a better description than
  an exponential function. We also find that the decay index is steep,
  1.8 on average, and different for every burst. This may be explained
  by contributions from degenerate electrons and photons to the
  specific heat capacity of the ignited layer and by deviations from
  the Stefan-Boltzmann law due to changes in the opacity with density
  and temperature. Detailed verification of this explanation yields
  inconclusive results. While the values for the decay index are
  consistent, changes of it with the burst time scale, as a proxy of
  ignition depth, and with time are not supported by model
  calculations.

\keywords{X-rays: binaries -- X-rays: bursts -- stars: neutron -- dense matter}}

\maketitle

%\linenumbers

\section{Introduction}
\label{sec:intro}

A common phenomenon among mass-transferring low-mass X-ray binaries
(LMXBs) with a neutron star (NS) as accretor is a thermonuclear shell
flash in the outer layer of that NS \cite[for reviews,
  see][]{lew93,stroh06}.  The matter accreted from the companion star
is rich in hydrogen and/or helium. It accumulates on the NS in a pile
thick enough that, at the bottom, a pressure is reached that is
sufficiently high for the ignition of thermonuclear fusion through the
CNO cycle and/or triple-$\alpha$ process. The ignition column depth is
$y=$10$^{8-12}$~g~cm$^{-2}$, while the geometric depth is
$10^{2-4}$~cm, compared to a NS radius of 10$^6$~cm. Often, the fusion
is unstable and most of the fuel is consumed within a fraction of a
second \cite[e.g.,][]{fuj81,bil98}. Temperatures reach values in
excess of 10$^9$~K at the location of ignition. Some of the heat is
conducted inward, but most is radiatively transported outward.  The
photosphere reaches temperatures of order $10^7$~K. The thermal
emission peaks in the X-ray regime of the spectrum, yielding a 'type
I' (thermonuclear) X-ray burst. Subsequently, the photosphere cools
down on a time scale determined by the amount of mass heated up. The
deeper the ignition is, the larger the mass heated up and the longer
the burst.

It is custom in the literature to model the decay phase of X-ray burst
light curves with an exponential function, the light curve being
defined as the time history of the number of photons that is detected
per unit time \citep[e.g.,][]{lew93,gal08}. The light curve in terms
of energy flux is in principle different, because the spectrum changes
during the decay as the NS cools. Therefore, the decay rate is not
necessarily the same. The difference is not dramatic, though. For the
bright phase of X-ray bursts, when the temperature is above 1 keV, the
peak of the energy ('$\nu F_\nu$') spectrum is above 3 keV which is for
a large part ($>80$\%) covered by the bandpass of most X-ray detectors
used thus far (2-30 keV).

While exponential decays are generally a good description of the X-ray
burst data, this is not based on physical considerations. A simplified
derivation of the expected decay law is as follows.  Let us assume
that the cooling layer has total mass $m$, temperature $T$ and a
specific heat capacity at constant pressure of $C_{\rm P}$. Then, the
amount of heat is given by
\begin{eqnarray}
Q & = & m \; C_P \; T
\label{eqn01}
\end{eqnarray}
If the heat is lost by radiation through a constant area $A$, the rate
of loss as a function of time $t$ is given by the Stefan-Boltzmann
law:
\begin{eqnarray}
\frac{{\rm d}Q}{{\rm d}t} & = & -A \sigma_{\rm sb} T_{\rm eff}^4 = -A \sigma_{\rm sb} {T^4\over \tau}
\label{eqn02}
\end{eqnarray}
where $\sigma_{\rm sb}$ the Stefan-Boltzmann constant and we assume
the layer has optical depth $\tau\gg 1$ with $\tau$ independent of
temperature.
%=5.67 \times 10^{-5}$ erg~cm$^{-2}$K$^{-4}$s$^{-1}$.
If heat is promptly transported within the reservoir before being
radiated through the photosphere and if the specific heat capacity is
independent of temperature, then
\begin{eqnarray}
 T & = & \left( \frac{3 A \sigma_{\rm sb}}{m\;C_P\tau} \, t \right)^{-1/3}
\label{eqn1}
\end{eqnarray}
and
\begin{eqnarray}
 L & = & -\frac{{\rm d}Q}{{\rm d}t} \propto T^4 \propto t^{-4/3}
\label{eqn2}
\end{eqnarray}
with $L$ the bolometric luminosity. Thus, the decay follows a power
law\footnote{An exponential decay function does apply in another
  circumstance: when the cooling is not due to radiation but to
  conduction. d$Q$/d$t$ is then proportional to the temperature
  difference $\Delta T$ with the cold medium instead of $\Delta
  T^4$}. This relationship is unaffected by General Relativity effects
close to the neutron star surface.

A more sophisticated study of NS cooling was performed by
\cite{cum04}, for X-ray bursts with large ignition depths
(carbon-fueled 'superbursts' with durations of roughly half a day),
using a multizone model that takes into account the heat transport
inside the reservoir after the flash. This study predicts the same 4/3
power-law decay index for late times in the burst, after the cooling
wave from the photosphere reaches the ignition depth. We will see
later (\S\ref{sec:discussion}) that the decay index may be different
after shallower ignitions.

We decided to verify the theory by checking whether a power law is
more consistent with the decay of X-ray bursts than an exponential
function, both in photon count rate and energy flux (or $L$). Our
study focuses on 'ordinary' X-ray bursts with durations of a few
minutes or less, because those are much more abundant and provide
better test data than long bursts. In \S\ref{sec:data} we explain how
we selected and prepared the data for this test, in
\S\ref{sec:modeling} we present details of the analysis method and
results, and in \S\ref{sec:discussion} these results are discussed.

\section{Data}
\label{sec:data}

\subsection{Observations}
\label{sec:data1}

The best data currently available are those collected with the
Proportional Counter Array \citep[PCA;][]{jah06} on the Rossi X-ray
Timing Explorer \citep[RXTE;][]{bra93} between 1996 and 2012. The PCA
consists of 5 Proportional Counter Units (PCUs) with a spectral
resolution of 1 keV at 6 keV (full-width at half maximum), a bandpass
of 2 to 60 keV and a combined photon-collecting area of about 6500
cm$^2$. This implies a typical X-ray burst peak photon count rate of
$10^4$ s$^{-1}$ (for 5 active PCUs) - the highest for any historical
X-ray telescope. During the course of the mission the average number
of active PCUs decreased, so that in general peak count rates were
higher earlier on in the mission. The PCA could simultaneously be read
out in 6 different data modes. For our analysis, we depend on 'event
mode data' (few to tens of ms readout resolution, 64 channels between
2 and 60 keV, PCUs unresolved), standard-1 data (0.125 s resolution,
no energy resolution, PCUs resolved) and standard-2 data (16 s
resolution, 129 energy channels, PCUs resolved).

\cite{gal08} published a catalog of 1187 X-ray bursts detected with
the PCA up to 2006. A final catalog is being generated as part of the
{\em Multi-INstrument Burst ARchive} effort ('MINBAR') to create an
archive of more than 6000 type I X-ray bursts detected with the PCA
and instruments on BeppoSAX and INTEGRAL \citep{gal10a}\footnote{More
  information about MINBAR is provided at URL {\tt
    burst.sci.monash.edu/minbar}}. That final catalog covers the
complete mission and includes 2097 PCA-detected X-ray bursts plus two
superbursts from 59 low-mass X-ray binaries.  These burst
identifications are the starting point of our study.

\subsection{Data selection}

Care has to be taken to obtain an unimpeded clean view of the NS
cooling process, because the signal may easily be confused by
prolonged nuclear burning through the rp process
\citep[e.g.,][]{wal81,han84,woo04,fis08}, scattering or obscuration in
the accretion disk or accretion disk corona
\citep[e.g.,][]{zan11,bag13b} and a varying accretion rate
\citep[e.g.,][]{wor13}. Therefore, X-ray bursts have to be carefully
selected.  Light curves were generated of all 2099 bursts. These are
histories of the count rate of detected photons as a function of
time. PCA 'standard-1' data were employed for this purpose with a time
resolution of 1 s, combining the signal of all active PCUs and photon
energies.

\begin{figure}[t]
\includegraphics[width=\columnwidth,angle=0]{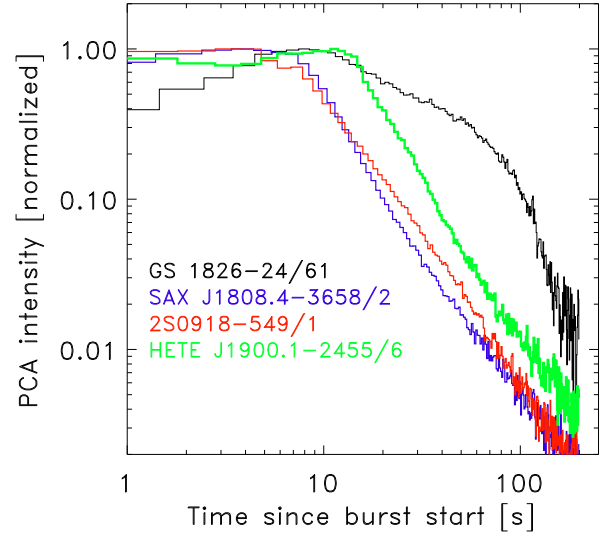}
\caption{Normalized light curves of three bursts of our sample (c.f.,
  Table~\ref{tab1}) and one from GS 1826-24 which has a strong
  additive component and whose tail cannot be due to cooling
  alone. The color version of this figure is only available online.
\label{fig:samen}}
\end{figure}

Visual inspection of the burst light curves resulted in the
identification of 14 different classes of decay shapes. We are
interested in bursts whose measurements are the least affected by
strong and variable non-burst emission or by variability that
indicates possible disturbances of the accretion flow
\citep[e.g.,][]{zan11}. Two classes contain bursts with the desired
shape of the decay: a smooth curve after a sharp peak or after a broad
less defined peak. Bursts in other classes show dents in the decay,
have other pre-burst fluxes than post-burst, show rise times similar
to their decay times or show multiple peaks without a quiescent period
in between.

The first class of smooth decays after sharp peaks is the largest with
655 bursts. That with smooth decays after broad peaks contains 119
bursts.  The total of 774 bursts encompass more than one third of all
RXTE bursts. Of the other classes, that with bursts with a shoulder
shape (see below) is largest (337 bursts). That with bursts with a
triangular shape is the least prominent with 7 bursts. It should be
noted that 405 bursts were weak and barely rose above the noise. The
many bursts from IGR J17480-2446 \citep{lin12} are good examples of
that.

To illustrate the difference between smooth and other decays, we show
in Fig.~\ref{fig:samen} a burst with a shoulder shape from GS~1826-24
together with 3 bursts that we ended up selecting.  Obviously, the
time profile of the burst from GS 1826-24 is rather different and is
neither consistent with a pure power law or an exponential
function. There is clearly an additional component which lasts 100 s
and then drops very fast. Such a burst tail cannot solely be due to
cooling. \cite{bil00} proposed that the shoulder in GS~1826-24 is due
to prolonged nuclear hydrogen burning through the relatively slow
rp-process. \cite{heg07a} convincingly proved this by detailed
modeling of the nuclear burning.

\begin{figure}[t]
\includegraphics[width=\columnwidth,angle=0]{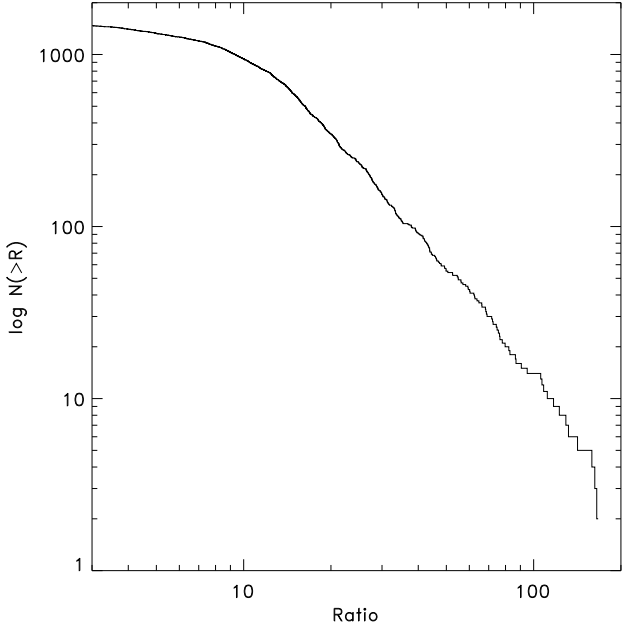}
\caption{Cumulative distribution of peak-to-pre-burst flux ratios
\label{fig:cumu}}
\end{figure}

Next, an additional selection criterion was applied. The best bursts
to study are those with the largest dynamic range in flux, the range
being defined as the ratio between the peak and the pre-burst fluxes,
because that avoids most of the confusion with changes in accretion
flux. The pre-burst flux was calculated by taking the average of the
flux in the 20 to 100~s time frame immediately prior to the burst
start time as determined by \cite{gal08}. The peak flux was determined
from the maximum between 10 s prior to 50 s posterior to the burst
start, at a time resolution of 1~s. The ratio $R$ is the dynamic
range. Fig. \ref{fig:cumu} shows the cumulative distribution of
$R$. To obtain a reasonably sized set of bursts with accurate enough
determinations of burst time scale parameters, we initially applied a
threshold of $R = 50$. This yields 22 bursts from the first class and
8 bursts from the second class (see above). However, seven bursts are
from the eclipsing system EXO~0748-676 and were removed because of a
high likelihood for interference by the accretion disk due to the high
inclination angle \citep{par86}. Furthermore, we had to leave out 4
bursts for which no event mode or burst catcher data are
available. This selection step, going from 755 to 19 bursts is the
most restrictive. We note that this does not introduce a selection
effect on shape.

To extend the diversity of NSs and bursts, we added bursts with
smaller $R$ as well as two long bursts. For the first addition, we
searched for LMXBs that had bursts with $R>10$ and picked 2 bursts per
LMXBs that were as far apart in time as possible to probe different
circumstances. This yielded 14 more bursts from 7 LMXBs. For the
addition of long bursts, there is not much choice in the PCA sample (4
bursts). We added an intermediate duration burst from 2S 0918-549
\citep{zan11} and a superburst from 4U 1636-536 \citep[][Keek et
  al. in prep.]{stroh02b,kuu04,cum06}. These two bursts do not have
monotonic decays and have low $R$, see \S\ref{sec:preparation}, but
particularly the superburst from 4U 1636-536 has the best data
available for such a long event.

Our total burst sample consists of 35 ordinary and 2 long bursts from
14 LMXBs, see Table~\ref{tab1}.  This includes a variety of LMXBs.
There are 3 confirmed ultracompact X-ray binaries with presumably a
deficiency in hydrogen (4U 0512-40, 2S 0918-549 and 4U 1820-30), 3
accretion-powered X-ray pulsars (IGR J17511-3057, SAX J1808.4-3658 and
HETE J1900.1-2455) and 6 transients (4U 1608-52, IGR J17511-3057, SAX
J1808.4-3658, XTE J1810-189, HETE J1900.1-2455 and Aql X-1). The rise
time of all 37 bursts is fast: the time to rise to 75\% of the burst
peak count rate is always less than 2 s. This automatically selects
flashes of 'pure' helium layers. Such layers exist either in a H-rich
LMXB when the accretion rate is in a favorable regime
\citep[e.g.,][]{fuj81} or in a H-poor LMXB in an ultracompact X-ray
binary system with a hydrogen-deficient companion/donor star. 29 out
of the 37 bursts are Eddington-limited (see Table~\ref{tab1}).

\subsection{Data preparation}
\label{sec:preparation}

For each burst we prepared two types of light curves. The first is the
history of the photon count rate in the detector. This is the same
kind of data that was used for the above data selection. We subtracted
for each burst the count rate as determined in a time frame of 20 to
100 s prior to the burst start time.

The second type of light curve is the history of the bolometric
flux. This involved a more elaborate data treatment. We employed
event-mode data, again combining all active PCUs, but resolved in
photon energy. In a few bursts (from 4U 1608-52, 4U 1728-34 and SAX
J1808.4-3658) the on-board buffer sometimes overflowed resulting in
data stretches not being downloaded and lost. That always happened
prior to the cooling phase and does not affect our analysis. We
selected calibrated data between 3 and 20 keV that are usually covered
by 23 energy channels. First, we generated a spectrum from pre-burst
standard-2 data as far as possible (up to 2500 s) and as far as it is
anticipated to be representative for the non-burst radiation during
the burst (i.e., with a flux that is identical within the noise to the
flux immediately prior to the burst). This spectrum was fitted, in
{\tt Xspec} \citep{arn96} version 12.8.0d, with a combination of a
disk black body \citep[e.g.,][]{mit84} and a power law, absorbed
following the model by \cite{mor83} for cosmic abundances and with
hydrogen column densities $N_{\rm H}$ fixed at values obtained from
the literature for each source \citep[see Appendix A in ][]{wor13}. A
systematic error of 0.5\% was added quadratically to the statistical
error per channel. For the vast majority of spectra this results in an
acceptable goodness of fit as measured through $\chi^2_\nu$. In
incidental cases $\chi^2_\nu$ was formally not acceptable but the
effect of that in our analysis was found to be negligible due to the
dominance of the burst flux over the persistent flux. Second, the
burst was divided in a number of time intervals for which separate
spectra were generated from event mode data. These were modeled with a
combination of the model for the pre-burst data and a Planck function
with a temperature $T{\rm c}$ and a normalization $R^2_{\rm 10~kpc}$
where $R_{\rm 10~kpc}$ is the radius of an assumed spherical emission
area in km for a distance of 10 kpc. All burst spectra were again
multiplied with the same absorption model and fixed hydrogen column
density $N_{\rm H}$ as above. Subsequently, the bolometric flux $F$
per burst time interval was determined through the law of
Stefan-Boltzmann:
\begin{eqnarray}
F(t) = \sigma_{\rm sb} 4\pi R^2_{\rm 10~kpc}/(10~{\rm kpc})^2 T{\rm
  c}^4~\hspace{1cm}{\rm erg~cm}^{-2}~{\rm s}^{-1},
\end{eqnarray}
where it is assumed that the emitting area is constant.  Statistical
errors for the bolometric flux were calculated through the same law,
by sampling parameter space 10,000 times in (-4$\sigma$,+4$\sigma$)
intervals ($\sigma$ representing the single-parameter 1-sigma error)
around the best-fit values of k$T_{\rm c}$ and $R^2_{\rm 10~kpc}$,
searching for all parameter pairs for which $\chi^2<\chi^2_{\rm
  min}+2.3$, calculating for those pairs the bolometric flux and
searching the minimum and maximum flux values for that $\chi^2$
constraint. These delimit the 68\% error margin in flux for two free
parameters \citep[e.g.,][]{lampton76}.

The two long bursts (2S 0918-549/5 and 4U 1636-536/sb) involve
additional data preparation. Both bursts do not have monotonic
decays. The intermediate duration burst from 2S 0918-549 has strong
modulations on its decay which extend from 105 to 201 s after the
start of the burst \citep{zan11}. We excluded data for this time
frame, leaving a few data points between 100 and 105 and between 201
and 226 s. The superburst from 4U 1636-536 has a very low $R$ value of
5.2 and we are forced to exclude a large part of the
tail. Furthermore, the cooling wave takes a long time to reach the
ignition depth, implying that it is necessary to exclude the first
3000 s of the burst. The left-over data covers 4708 to 8616 s after
the start of the burst, compared to a data set extending 20,000 s
(including data gaps).

Instrumental dead time corrections were applied to both the bolometric
flux values and the photon count rate values.

\section{Light curve modeling}
\label{sec:modeling}

We tested two models for the evolution of the flux during the X-ray
burst decay. The first is the traditional exponential function:
\begin{eqnarray}
F(t) & = & F_0 \;\; e^{-(t-t_0) / \tau} + F_{\rm bg},
\end{eqnarray}
and the second the power law function
\begin{eqnarray}
F(t) & = & F_0 \left( \frac{t-t_s}{t_0-t_s} \right)^{-\alpha} + F_{\rm
  bg}
\end{eqnarray}
where $t$ is time, $t_0$ the time where $F_0$ is measured and $t_s$
the time when the cooling starts (irrelevant for the exponential
function). $\tau$ is the exponential decay folding time and $\alpha$
the power-law decay index. $F_{\rm bg}$ is the background flux (i.e.,
everything unrelated to the burst emission and assumed to be
constant). It is fixed at 0 for all fits (but see below). $t_0$ is
chosen to be the time of the first data point included in the fit. The
typical burst time scale is $\tau$ for the exponential function and
$t_0-t_s$ for the power law, if $t_0$ is the time when the decay
starts.

\begin{table*}[t]
\begin{center}
\caption[]{Fit results on 35 ordinary bursts and 2 long bursts (at the
  bottom, below the line). The bursts numbers (column 2) are those
  from the burst catalog partly published in
  \cite{gal08}. Uncertainties are only provided for $\alpha$ since
  only power-law fits are acceptable. Sometimes 2 values for
  $\chi^2_\nu$ are provided. The fitted values for $\alpha$ apply to
  the low value of $\chi^2_\nu$ (see text).\label{tab1}}
\begin{tabular}{|lrcr|rrrr|rrrr|}
\hline\hline
Object & Bu. & \multicolumn{1}{c}{MJD} & $R$ & \multicolumn{4}{c|}{Photon count rate history} & \multicolumn{4}{c|}{Bol. flux history}\\
       &  No.  & \multicolumn{1}{c}{}    &     & \multicolumn{2}{c|}{Exponential} & \multicolumn{2}{c|}{Power law} & \multicolumn{2}{c|}{Exponential} & \multicolumn{2}{c|}{Power law} \\
       &       & \multicolumn{1}{c}{}    &     & \multicolumn{1}{c}{$\tau$} & \multicolumn{1}{c|}{$\chi^2_\nu$} & \multicolumn{1}{c}{$\alpha$} & \multicolumn{1}{r|}{$\chi^2_\nu$} & \multicolumn{1}{c}{$\tau$} & \multicolumn{1}{c|}{$\chi^2_\nu$} & \multicolumn{1}{c}{$\alpha$} & \multicolumn{1}{c|}{$\chi^2_\nu$} \\
\hline
4U 0512-40*       &   2 & 51324.286947 &  22.3 & 13.8 &  3.0 & $1.707\pm0.018$  & 1.15 & $ 7.9$ & 14.3 & $1.439\pm0.022$ & 0.38 \\
4U 0512-40*       &  15 & 54839.516222 &  27.7 & 11.4 &  2.5 & $1.715\pm0.024$  & 0.97 & $ 7.7$ &  7.1 & $1.792\pm0.042$ & 0.65 \\
2S 0918-549       &   1 & 51676.826588 & 122.6 & 19.2 & 17.7 & $1.902\pm0.007$  & 1.41 & $13.8$ & 65.9 & $1.832\pm0.013$ & 1.69 \\
4U 1608-52        &   8 & 50914.274663 &  86.9 & 14.7 & 90.2 & $1.894\pm0.004$  & 1.08 & $14.5$& 114.9 & $1.808\pm0.016$ & 1.74/8.31 \\
4U 1608-52        &   9 & 51612.030846 &  59.5 & 10.1 & 63.2 & $1.829\pm0.004$  & 1.49 & $ 3.8$ &  1.2 & $2.141\pm0.049$ & 2.45 \\
4U 1608-52        &  10 & 51614.071258 & 111.5 & 14.0 &103.5 & $1.995\pm0.003$  & 0.95 & $ 9.9$ &248.8 & $2.000\pm0.013$ & 1.05/8.86 \\
4U 1608-52        &  31 & 53104.407932 &  90.7 & 17.8 & 64.5 & $1.859\pm0.004$  & 5.14 & $13.0$ &256.9 & $1.869\pm0.015$ & 2.13/9.52 \\
4U 1636-536       &  68 & 52286.054034 &  35.5 &  9.2 & 10.2 & $1.681\pm0.010$  & 1.34 & $ 4.9$ & 82.6 & $1.652\pm0.011$ & 2.45 \\
4U 1636-536       & 327 & 55394.904042 &  31.2 & 12.1 &  4.9 & $1.681\pm0.016$  & 1.06 & $ 5.2$ & 43.7 & $1.592\pm0.014$ & 1.67 \\
4U 1702-429*      &  13 & 51939.193940 &  42.9 & 12.5 & 12.6 & $1.816\pm0.009$  & 1.17 & $ 6.1$ & 82.8 & $1.751\pm0.009$ & 2.02 \\
4U 1702-429       &  44 & 53212.793589 &  43.6 & 12.0 & 33.4 & $1.825\pm0.006$  & 1.43 & $ 7.8$ &106.2 & $1.874\pm0.012$ & 1.81/3.00 \\
4U 1705-44*       &  51 & 54046.201890 &  33.9 & 12.4 &  4.4 & $1.863\pm0.017$  & 1.46 & $ 9.4$ & 17.9 & $1.801\pm0.021$ & 0.79 \\
4U 1705-44*       &  77 & 55062.220583 &  29.8 & 11.4 &  2.7 & $1.807\pm0.025$  & 1.05 & $ 5.7$ & 19.7 & $1.421\pm0.017$ & 0.68 \\
4U 1724-30        &   2 & 53058.401400 &  16.9 & 11.4 &  3.0 & $1.823\pm0.028$  & 0.96 & $ 4.6$ & 28.5 & $1.651\pm0.018$ & 0.99 \\
4U 1724-30        &   3 & 53147.218284 &  27.8 & 12.9 & 11.5 & $1.764\pm0.014$  & 1.31 & $ 5.4$ & 71.1 & $1.857\pm0.012$ & 1.19 \\
4U 1728-34        &  76 & 51657.203264 &  33.0 &  8.1 & 93.1 & $1.786\pm0.006$  & 2.19 & $ 6.7$ & 76.1 & $1.781\pm0.011$ & 1.18 \\
4U 1728-34        & 126 & 54120.25887  &  29.6 &  7.6 & 68.4 & $1.784\pm0.007$  & 2.02 & $ 6.4$ & 67.0 & $1.835\pm0.010$ & 1.00 \\
IGR J17511-3057*  &  10 & 55099.313613 &  43.3 & 12.4 &  3.9 & $2.303\pm0.026$  & 1.09 & $10.2$ & 14.2 & $2.320\pm0.033$ & 1.89 \\
IGR J17511-3057*  &  12 & 55101.289836 &  47.4 & 12.4 &  3.6 & $2.134\pm0.028$  & 1.40 & $ 9.8$ &  7.9 & $2.065\pm0.031$ & 2.05 \\
SAX J1808.4-3658  &   2 & 52564.305146 &  63.8 & 25.2 & 13.8 & $1.820\pm0.010$  & 1.57 & $20.4$ & 32.6 & $1.789\pm0.023$ & 1.27 \\
SAX J1808.4-3658  &   3 & 52565.184268 &  74.8 & 24.9 & 27.0 & $1.896\pm0.008$  & 2.39 & $22.0$ & 50.3 & $1.814\pm0.050$ & 0.88/6.50 \\
SAX J1808.4-3658  &   4 & 52566.426770 &  82.8 & 27.0 & 22.9 & $1.984\pm0.008$  & 1.93 & $17.0$ &113.1 & $1.961\pm0.033$ & 1.75/9.30 \\
SAX J1808.4-3658  &   6 & 53526.638240 &  76.5 & 29.1 & 18.6 & $1.868\pm0.010$  & 2.20 & $17.0$ & 71.6 & $1.954\pm0.048$ & 0.67/6.27   \\
SAX J1808.4-3658  &   7 & 54732.708128 &  95.0 & 30.2 &  8.2 & $2.089\pm0.017$  & 1.78 & $16.4$ & 53.3 & $2.231\pm0.041$ & 2.67/8.13 \\
SAX J1808.4-3658  &   9 & 55873.916348 &  79.7 & 25.3 & 24.1 & $1.903\pm0.008$  & 1.45 & $19.1$ & 40.8 & $1.739\pm0.029$ & 4.22/10.34 \\
SAX J1810.8-2609  &   3 & 54590.729819 &  62.5 & 11.2 & 13.6 & $1.833\pm0.010$  & 1.14 & $11.6$ &  9.2 & $1.633\pm0.029$ & 1.24 \\
4U 1820-30        &   5 & 53277.438562 &  13.3 &  6.4 &  5.4 & $2.002\pm0.016$  & 3.33 & $ 5.4$ & 25.5 & $1.991\pm0.020$ & 0.75 \\
4U 1820-30        &  12 & 54981.187286 &  15.1 &  5.2 & 14.3 & $1.901\pm0.011$  & 5.28 & $ 7.2$ & 27.0 & $1.885\pm0.021$ & 2.09 \\
HETE J1900.1-2455 &   3 & 54506.856149 &  56.1 & 11.4 & 30.3 & $2.155\pm0.007$  & 3.22 & $ 7.9$ &100.9 & $2.276\pm0.015$ & 1.74/3.01 \\
HETE J1900.1-2455 &   5 & 54925.796423 &  72.2 & 14.7 & 16.5 & $1.858\pm0.008$  & 1.31 & $ 9.8$ & 46.6 & $1.727\pm0.015$ & 1.77 \\
HETE J1900.1-2455 &   6 & 55384.878220 &  86.5 & 14.7 & 38.2 & $2.216\pm0.007$  & 2.99 & $ 9.3$ &137.4 & $2.404\pm0.015$ & 6.31/11.16 \\
HETE J1900.1-2455 &   7 & 55459.228637 &  59.3 & 11.3 & 50.1 & $2.234\pm0.005$  & 3.59 & $ 9.3$ &201.6 & $2.289\pm0.014$ & 3.71/11.09 \\
Aql X-1           &  11 & 51336.590743 &  64.8 & 11.0 & 25.8 & $1.802\pm0.007$  & 1.18 & $12.3$ & 14.2 & $1.531\pm0.027$ & 1.73 \\
Aql X-1           &  25 & 52100.799520 &  56.2 &  8.5 & 45.5 & $1.836\pm0.005$  & 1.25 & $14.5$ & 14.3 & $1.641\pm0.026$ & 1.01 \\
Aql X-1           &  64 & 54259.247877 & 162.5 & 24.6 & 67.7 & $2.078\pm0.004$  & 2.56 & $21.6$ &108.3 & $1.904\pm0.044$ & 0.26/6.64 \\
\hline
2S 0918-549       &   5 & 54504.126944 & 158.8 & 110.6 & 3.4 & $1.372\pm0.007$  & 1.79 & $99.3$ &  3.9 & $1.516\pm0.005$ &  1.47 \\
4U 1636-536*      &  sb & 51962.702961 &   5.2 &4387.1 & 1.5 & $1.428\pm0.004$  & 1.23 &$4879$  &  5.0 & $1.321\pm0.004$ &  1.59 \\
\hline\hline
\end{tabular}
\end{center}
* non-PRE burst
\end{table*}

Fitting the exponential function to the data involves finding the best
values for $\tau$, $F_0$ and $F_{\rm bg}$. Fitting the power law
involves 4 instead of 3 parameters: $\alpha$, $F_0$, $F_{\rm bg}$ and
$t_s$. In principle one expects $t_s$ to be close to the start time of
the burst. In both cases we determined $F_{\rm bg}$ from pre-burst
data, assuming that during the burst this is not different, see
\S\ref{sec:preparation}.

\begin{figure}[t]
\includegraphics[width=\columnwidth,angle=0]{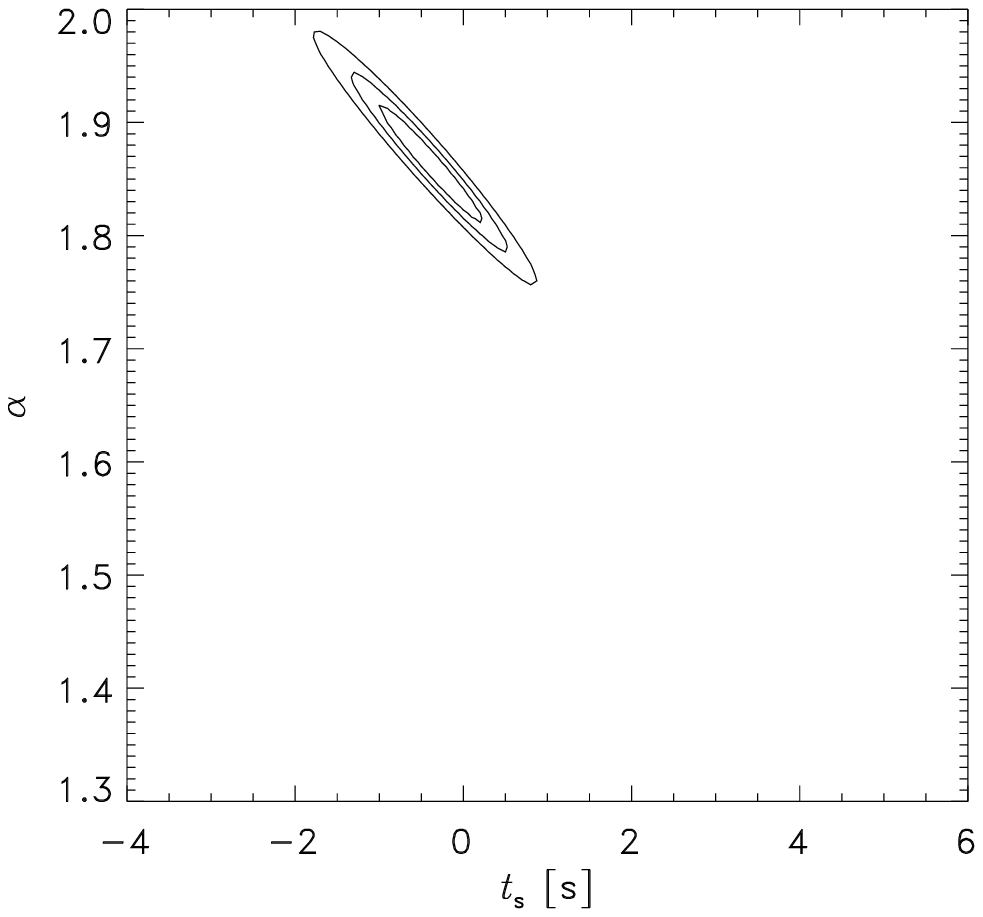}
\includegraphics[width=\columnwidth,angle=0]{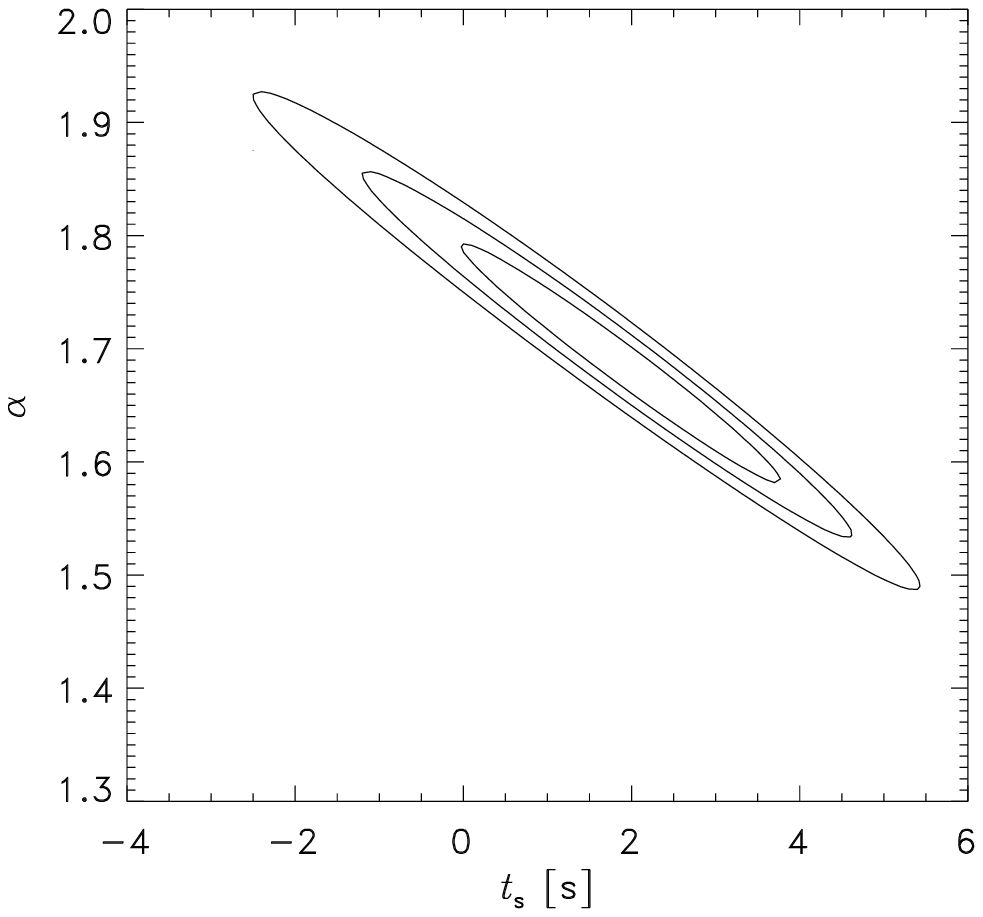}
\caption{Contours of goodness of fit $\chi^2_\nu$ for a power-law fit
  to the bolometric flux data for the burst from 0918-549 (upper plot)
  and burst number 2 from SAX J1808.4-3658 (lower) as a function of
  $t_{\rm s}$ and $\alpha$. Three contours are drawn, for $\chi^2_{\rm
    min}+\Delta \chi^2$ with $\Delta \chi^2=1.0, 2.3$ and 4.6 (i.e.,
  not per degree of freedom). The first contour delimits the single
  parameter 1$\sigma$ region. The second and third contour delimit the
  68\% and 90\% confidence regions. The contours show a coupling
  between both parameters.
\label{fig:chi}}
\end{figure}

There is a fundamental difference between both functions. The power
law is, for positive $\alpha$, divergent for $t=t_{\rm s}$ while the
exponential function has no divergence point. That causes a strong
coupling between $\alpha$ and $t_{\rm s}$
\citep[e.g.,][]{clau09}. Since $t_{\rm s}$ is outside the range of $t$
for which there are data, this may induce a large error on
$\alpha$. We work on the presumption that $t_{\rm s}$ is accurately
given by the burst start time, but have to keep in mind that on rare
occasions this may not be known. Some superexpansion bursts have
precursors that are quite short - of order tens of ms
\citep{zan10,zan11}. Fortunately, the PCA instrument is quite
sensitive and can pick up small signals, but if precursors are as
short as a few ms, that may even be a problem for the PCA. In that
case $t_{\rm s}$ may be off by order 1 s. We did vary $t_{\rm s}$ for
our bursts, to check whether better fits were possible for start times
very different from that of burst onset, but were unable to find such
instances (see Fig.~\ref{fig:chi} for such exercises on bursts from 2S
0918-549 and SAX J1808.4-3658).

It is necessary to skip the first part of the burst, because that is
not smoothly decaying yet. For each burst we increased $t_0$ from
immediately after the peak until $\chi^2_\nu$ did not decrease
anymore. This usually implies that the first 10 s of the burst,
including the rising part, are skipped and the flux decreased by
approximately a factor of 2. This ensures exclusion of that part of
the burst where possibly nuclear burning is ongoing or where the flux
is close to the Eddington limit during which part of the radiated
energy may be transformed to kinetic and potential energy instead of
radiation.

Similarly, it is sensible to not include all data beyond $t_0$ but to
stop when the burst flux becomes of the same order of magnitude as the
pre-burst flux. We did this for the fits to the bolometric flux
data. For the photon count rate data we mostly included all data until
200 s after burst onset and longer for the long bursts. Since the data
preparation is in this case more straightforward, we thought it
interesting to study the decay further down in the tail. This did not
result in more insight, though. Inconsistent power law fits are mostly
due to random changes in slopes (i.e., shallow to steep and vice
versa). We note that, for both sets of data, we used the same data for
the fits with the exponential function as for those with the power
law.

Table~\ref{tab1} presents the results of the modeling of the photon
count rate and bolometric flux data of the 37 selected bursts.
Comparing the goodness-of-fits $\chi^2_\nu$ of the exponential and
power-law fits, it is clear that power laws are, for every burst, a
better description of both the photon count rate and the bolometric
flux data.

In 13 bursts, fitting the power-law function to the bolometric fluxes
yields unacceptable values for the goodness-of-fit $\chi^2_\nu$,
although better than for the exponential model. That is probably due
to the fact that some spectra in those time series have large values
for $\chi^2_\nu$. In order to obtain more reliable estimates of the
uncertainty in the decay index, we multiplied in these cases the
errors on flux per time bin with $\sqrt{\chi^2_\nu}$ of the
appropriate spectral fit to force $\chi^2_\nu$ to 1 and performed the
power-law fit again and determine the uncertainty in $\alpha$. For
reference, the $\chi^2_\nu$ values before and after this procedure are
provided in Table~\ref{tab1} (last column). There is always
considerable improvement in $\chi^2_\nu$ in the power law fits. The
reason why the spectra are formally inconsistent with a black body is
unclear. It may be related to transient scattering effects in the
accretion flow. If so, then these bursts should, according to our
selection requirements not be included in our sample. Therefore, these
bursts should be considered with caution.

\begin{figure*}[p]
\includegraphics[width=0.42\columnwidth,angle=0]{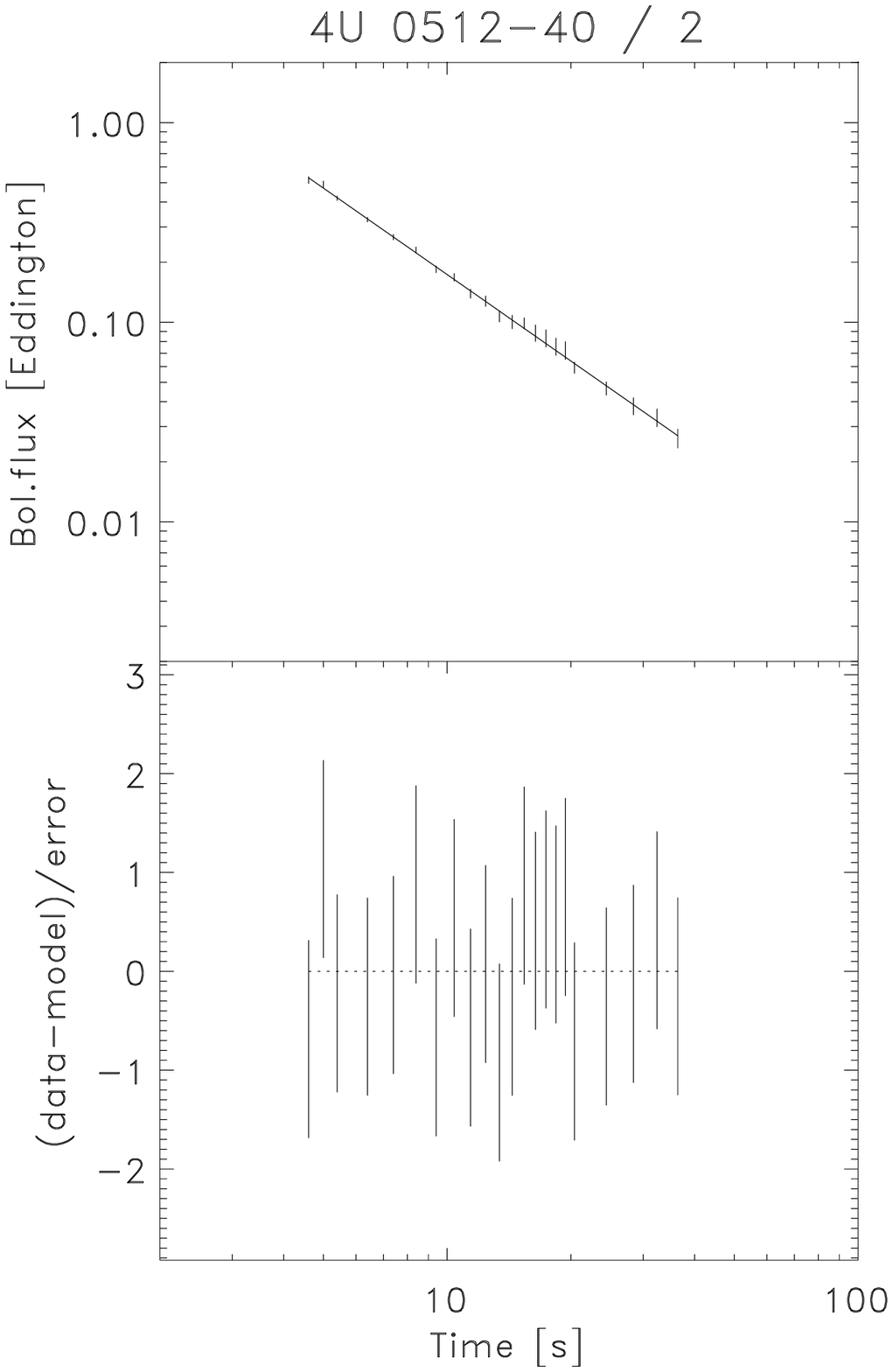}
\includegraphics[width=0.42\columnwidth,angle=0]{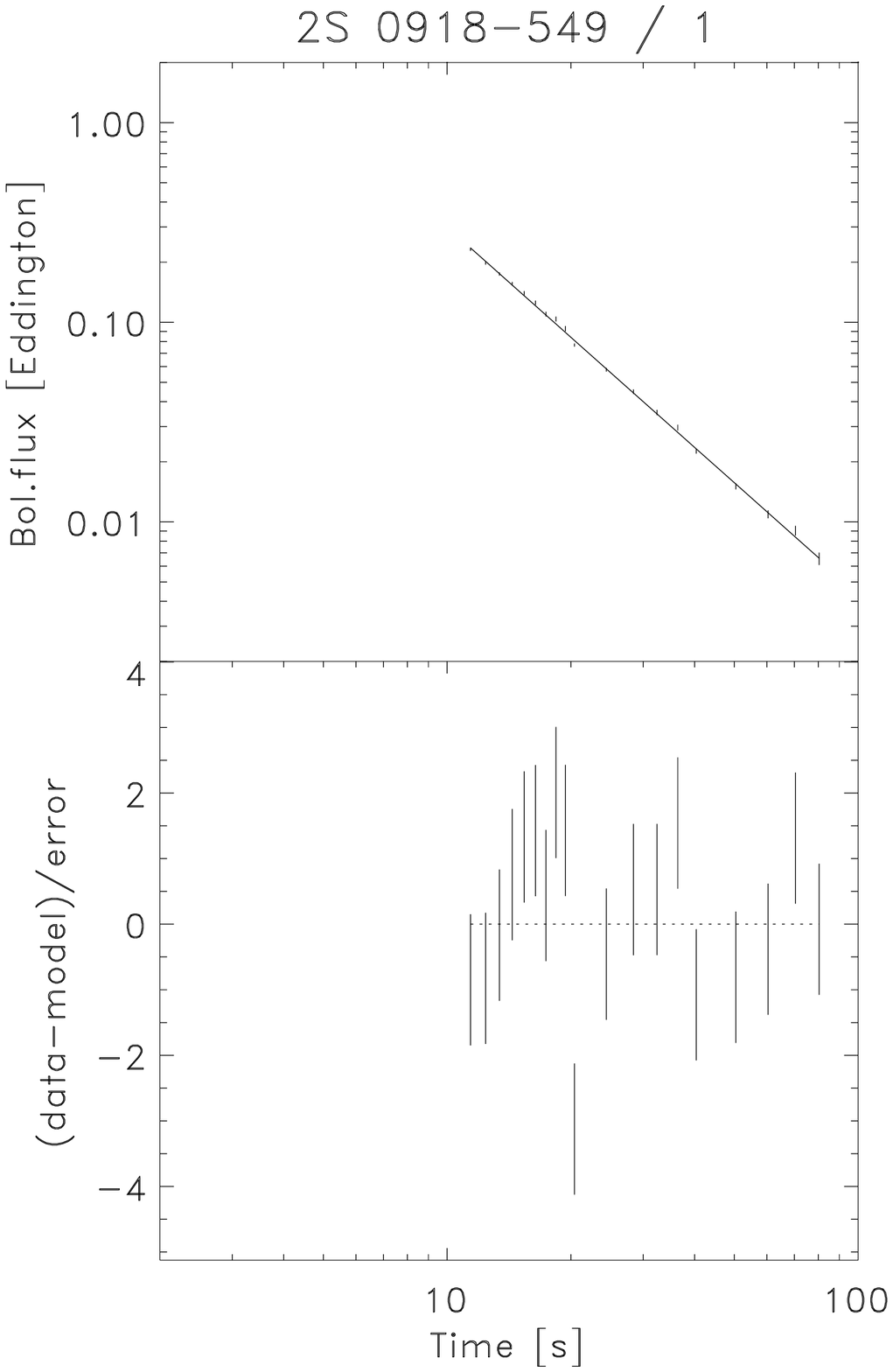}
\includegraphics[width=0.42\columnwidth,angle=0]{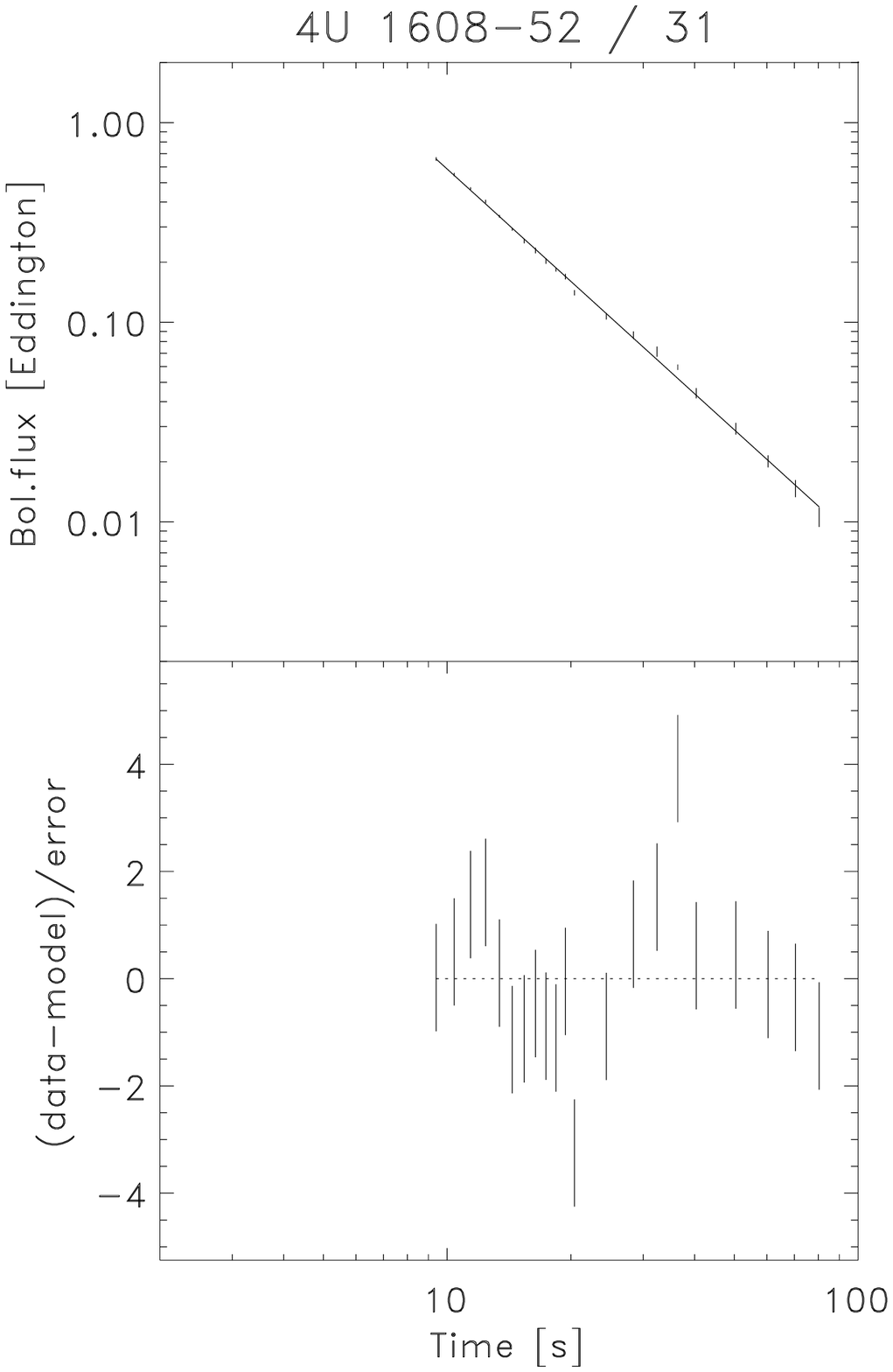}
\includegraphics[width=0.42\columnwidth,angle=0]{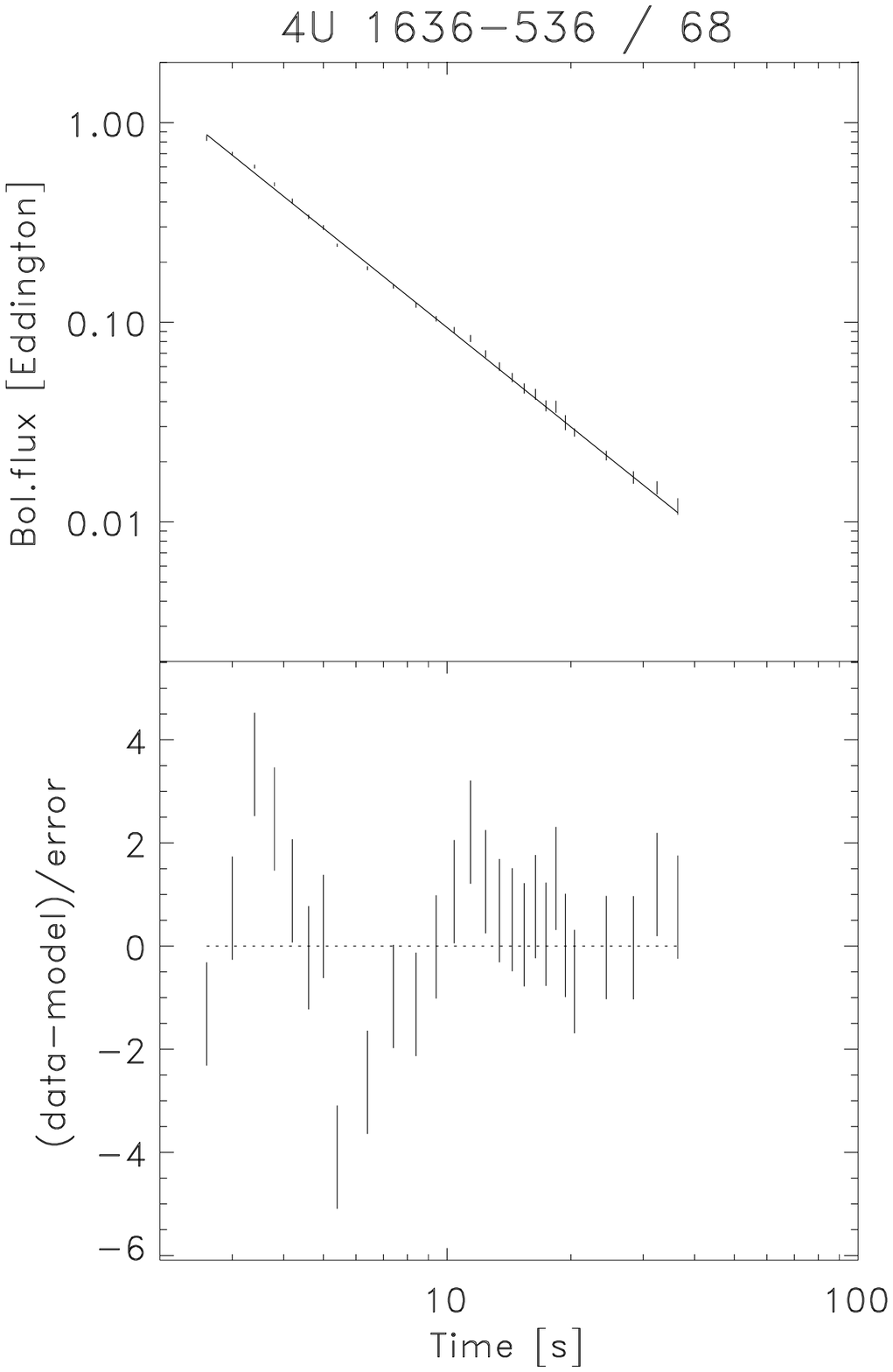}
\includegraphics[width=0.42\columnwidth,angle=0]{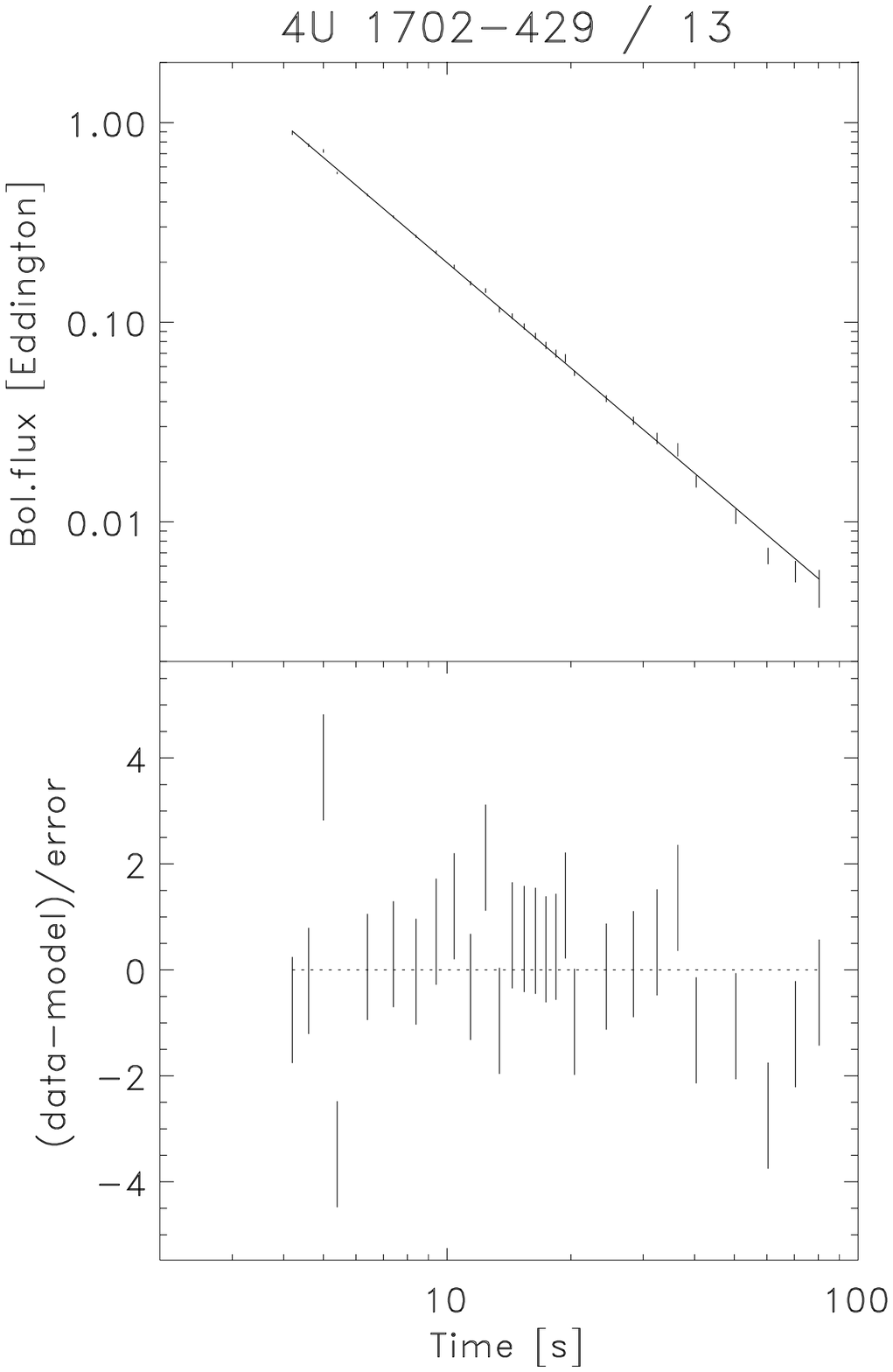}
\includegraphics[width=0.42\columnwidth,angle=0]{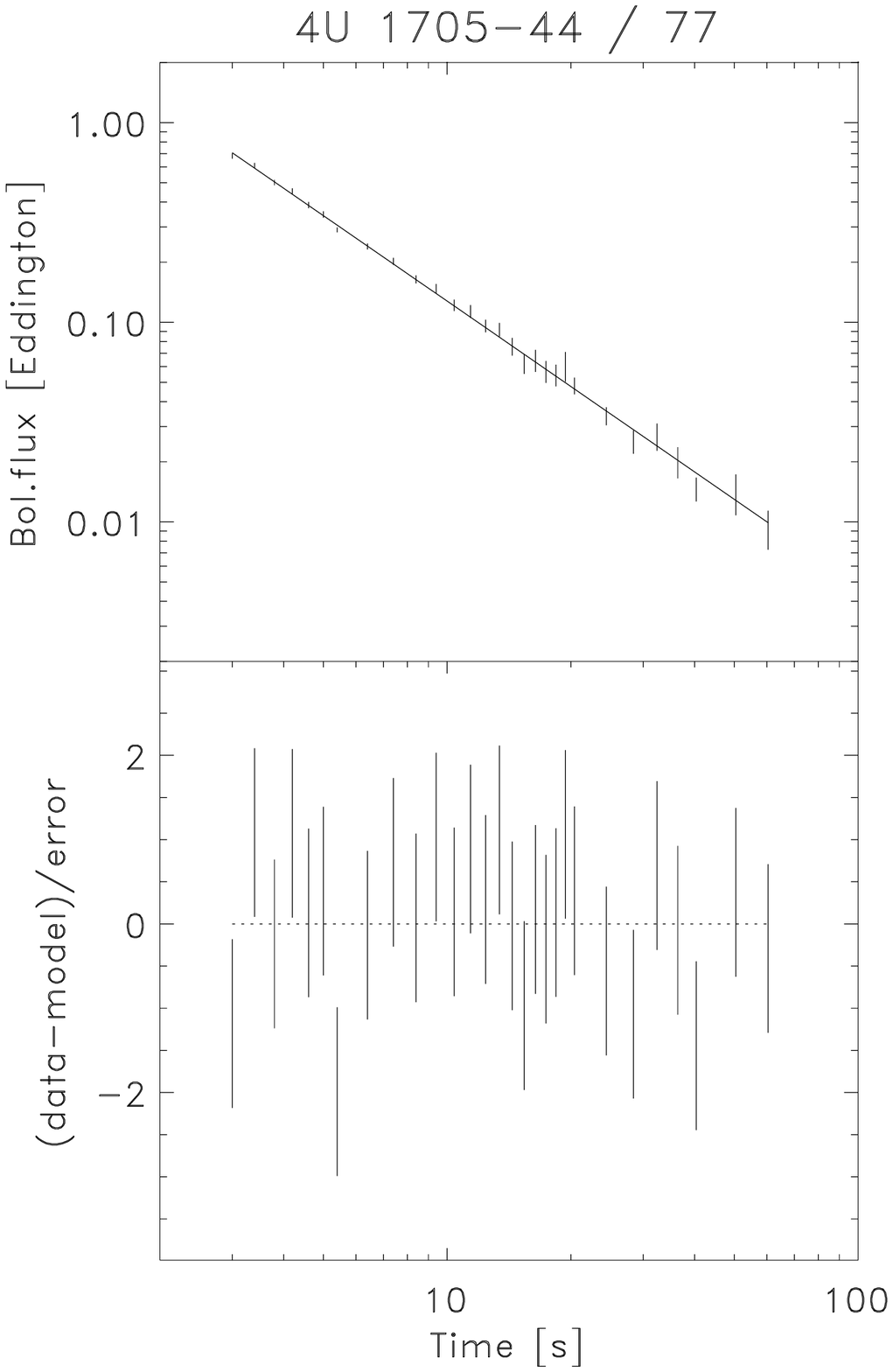}
\includegraphics[width=0.42\columnwidth,angle=0]{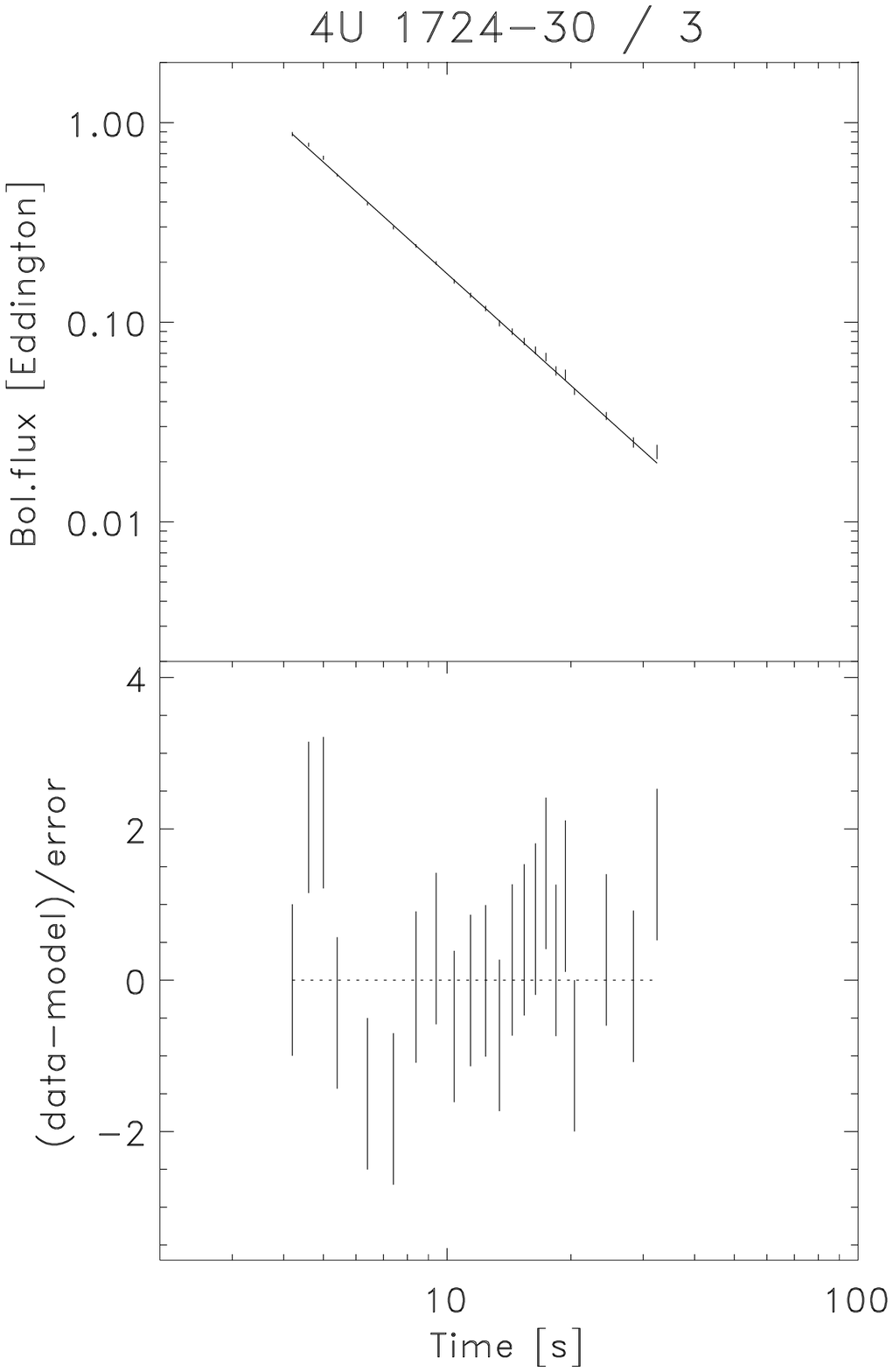}
\includegraphics[width=0.42\columnwidth,angle=0]{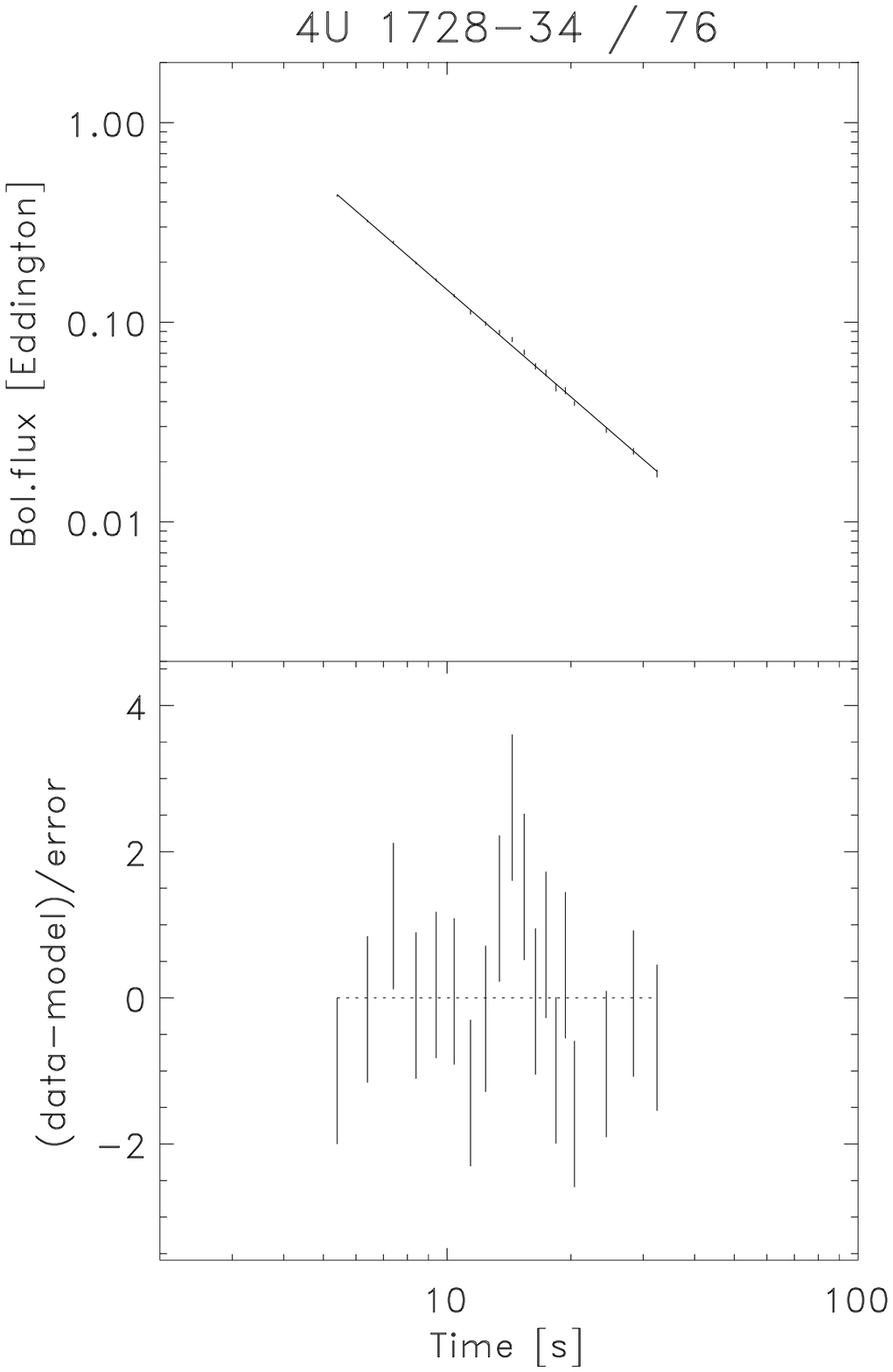}
\includegraphics[width=0.42\columnwidth,angle=0]{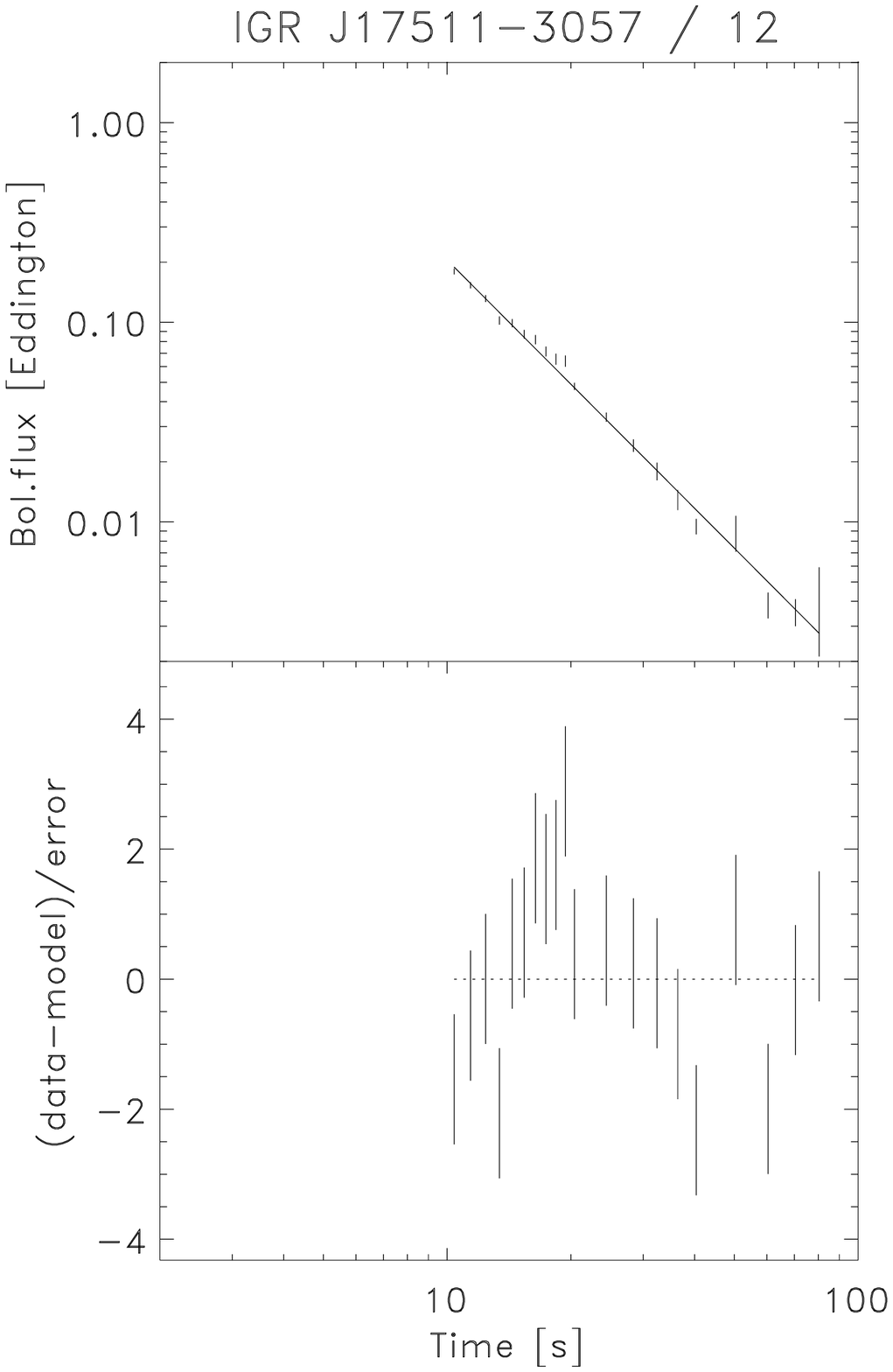}
\includegraphics[width=0.42\columnwidth,angle=0]{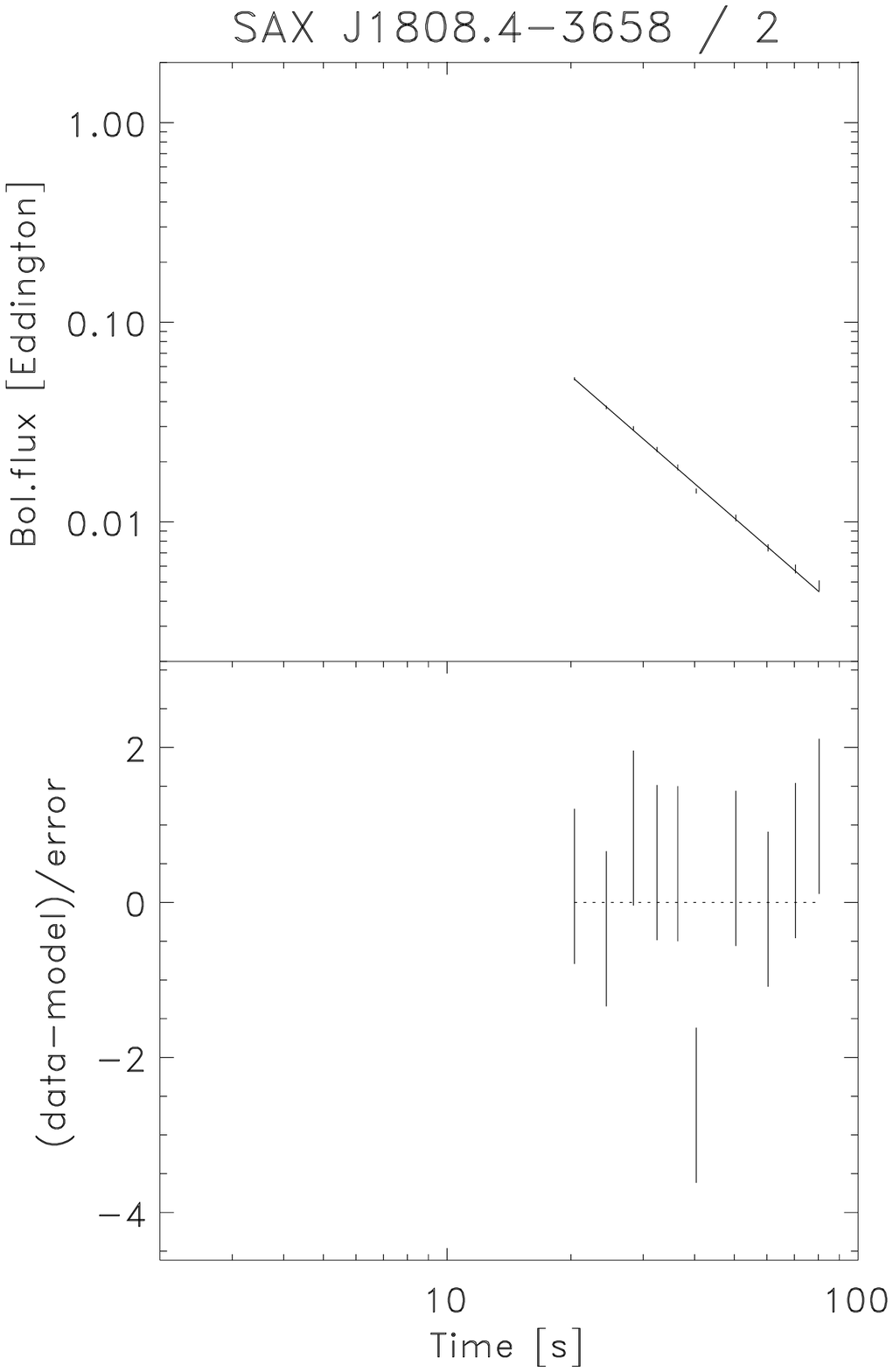}
\includegraphics[width=0.42\columnwidth,angle=0]{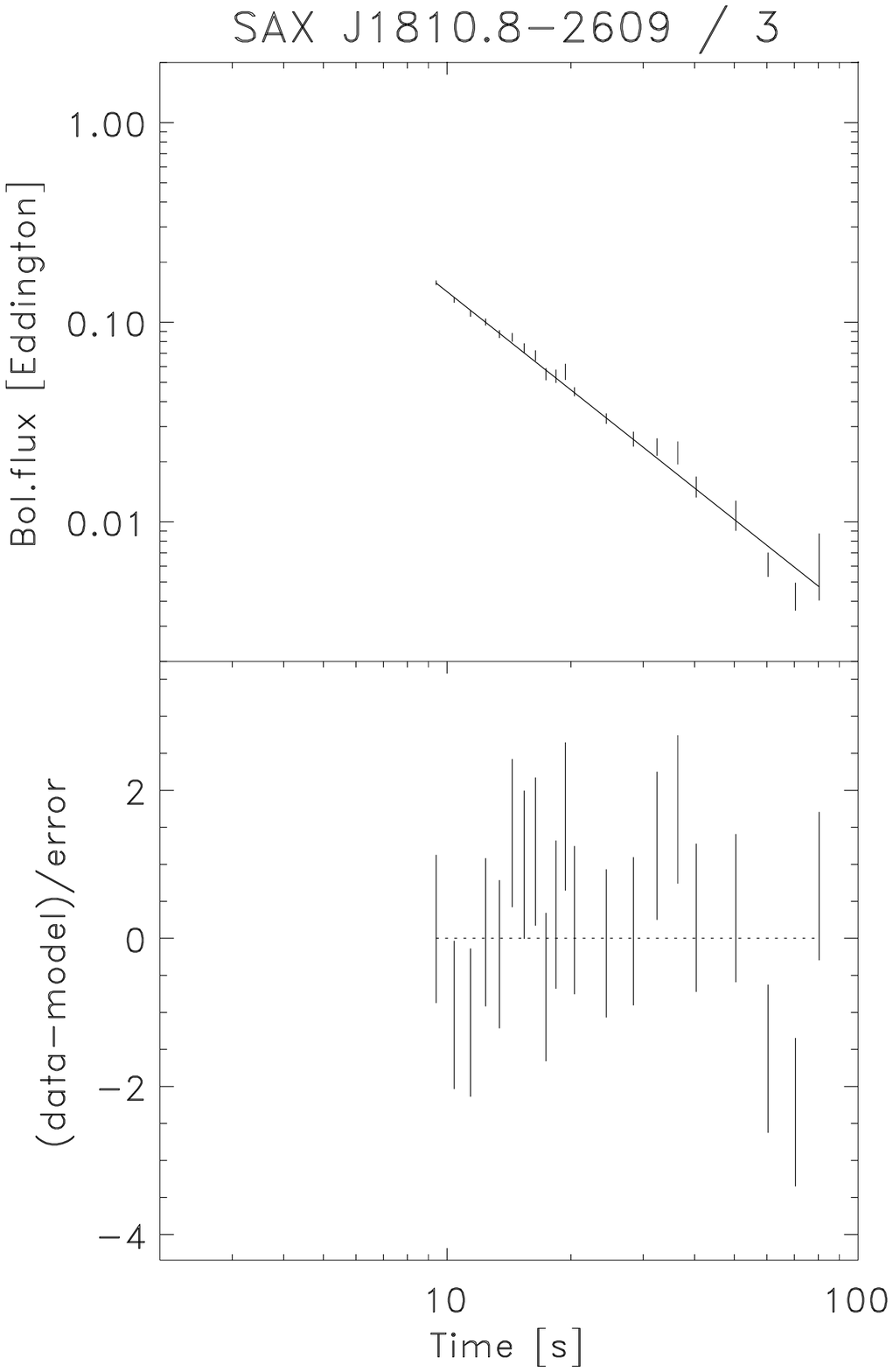}
\includegraphics[width=0.42\columnwidth,angle=0]{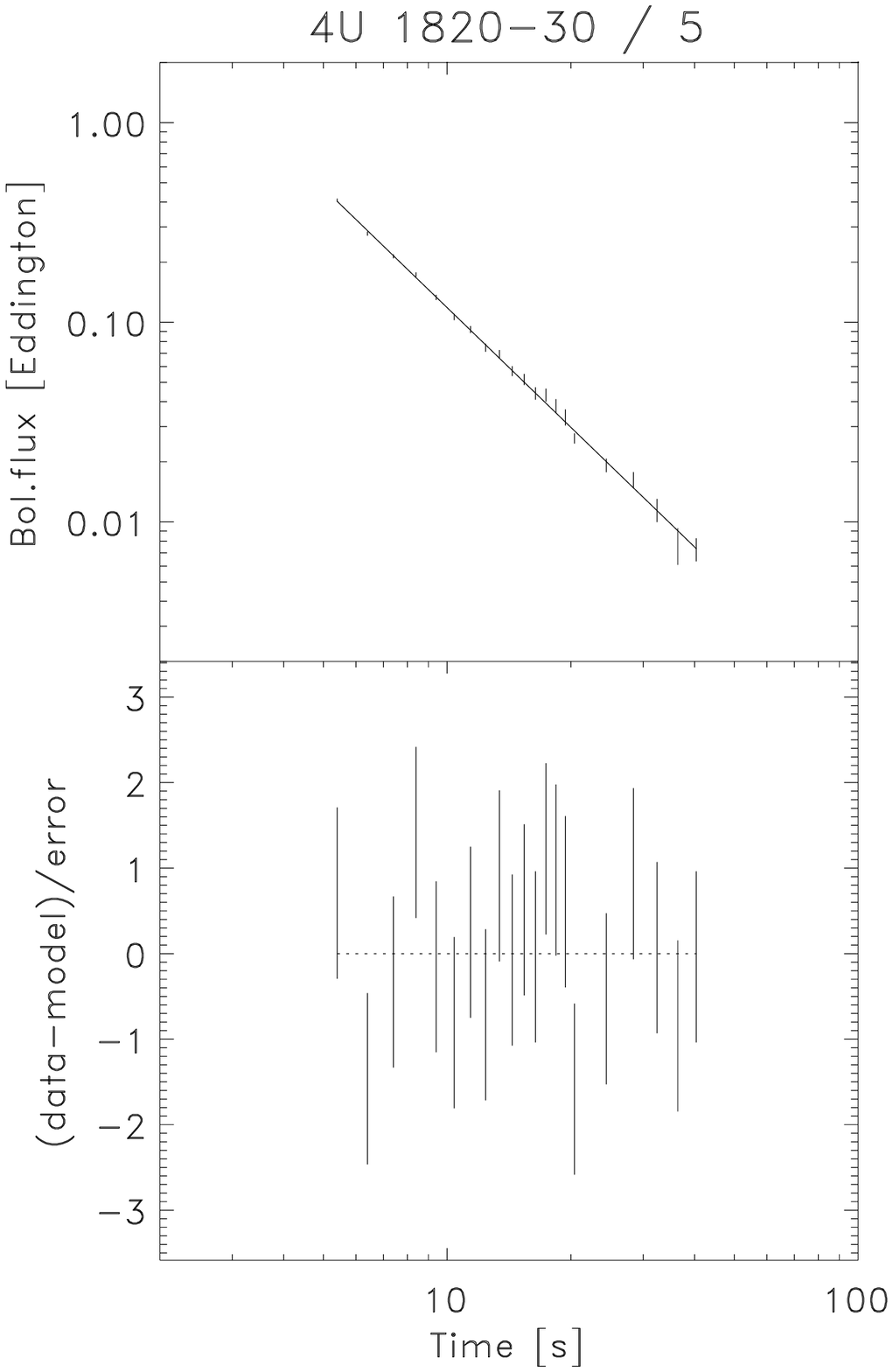}
\includegraphics[width=0.42\columnwidth,angle=0]{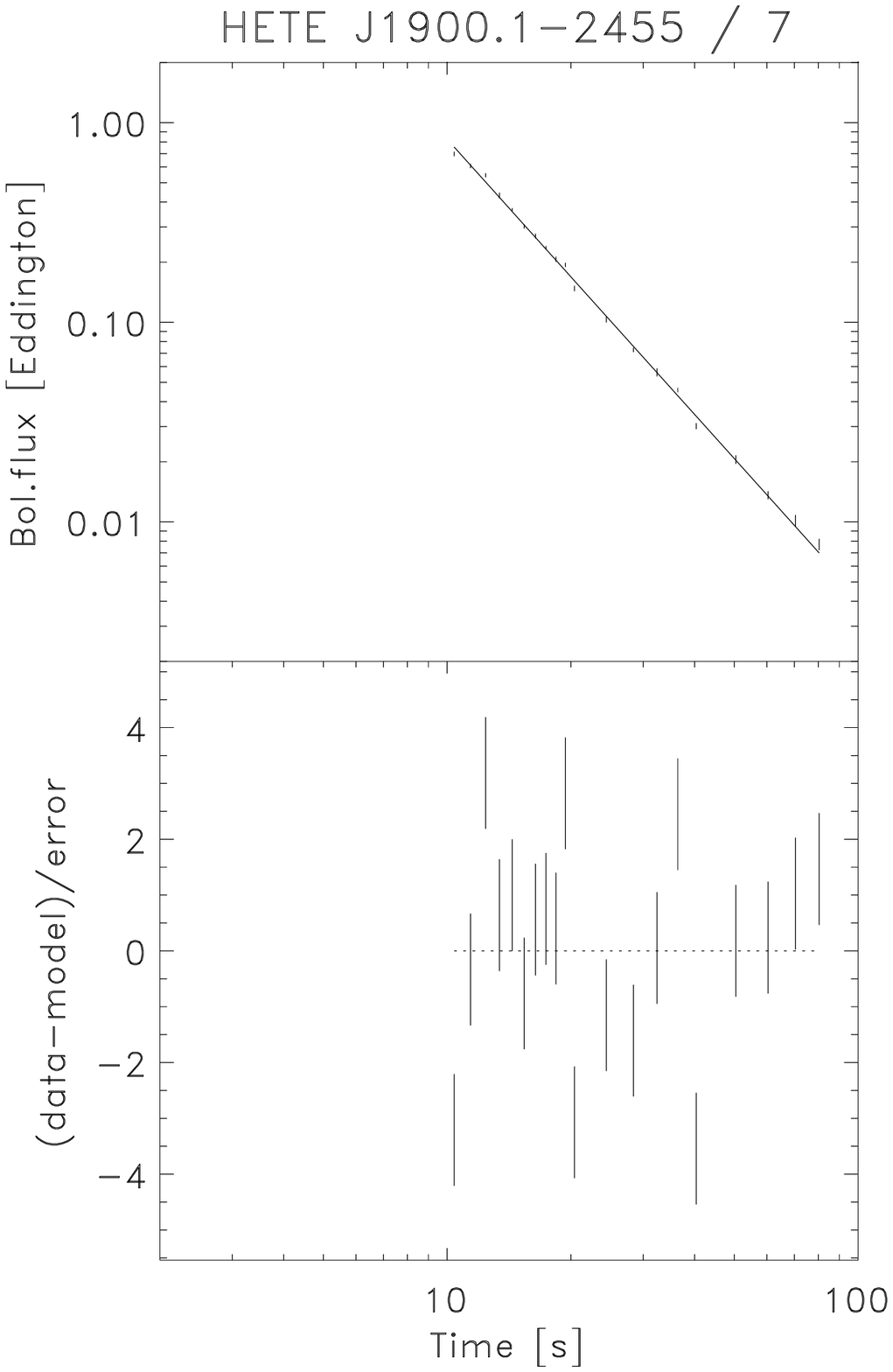}\hspace{10mm}
\includegraphics[width=0.42\columnwidth,angle=0]{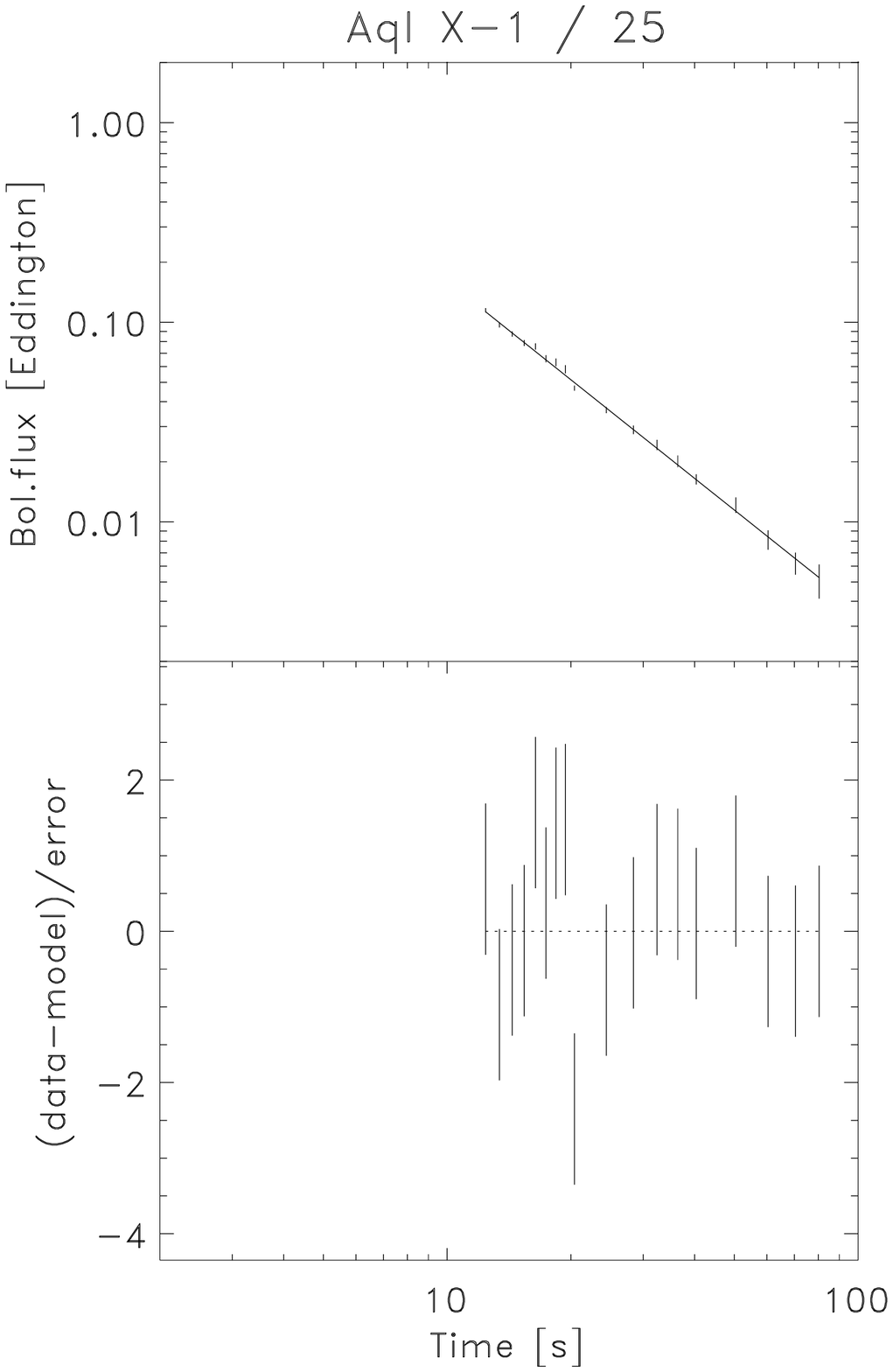}\hspace{10mm}
\includegraphics[width=0.42\columnwidth,angle=0]{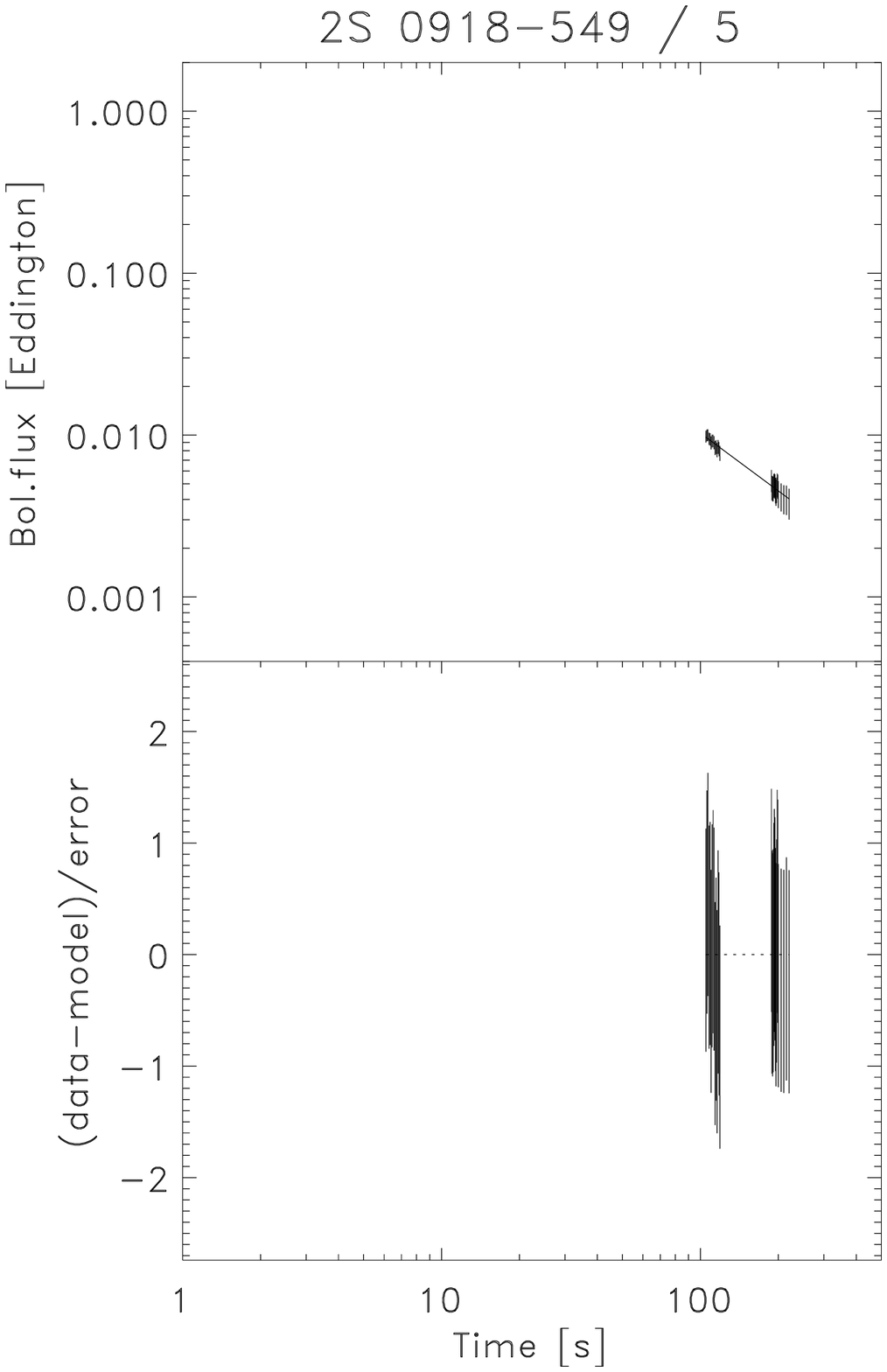}\hspace{10mm}
\includegraphics[width=0.42\columnwidth,angle=0]{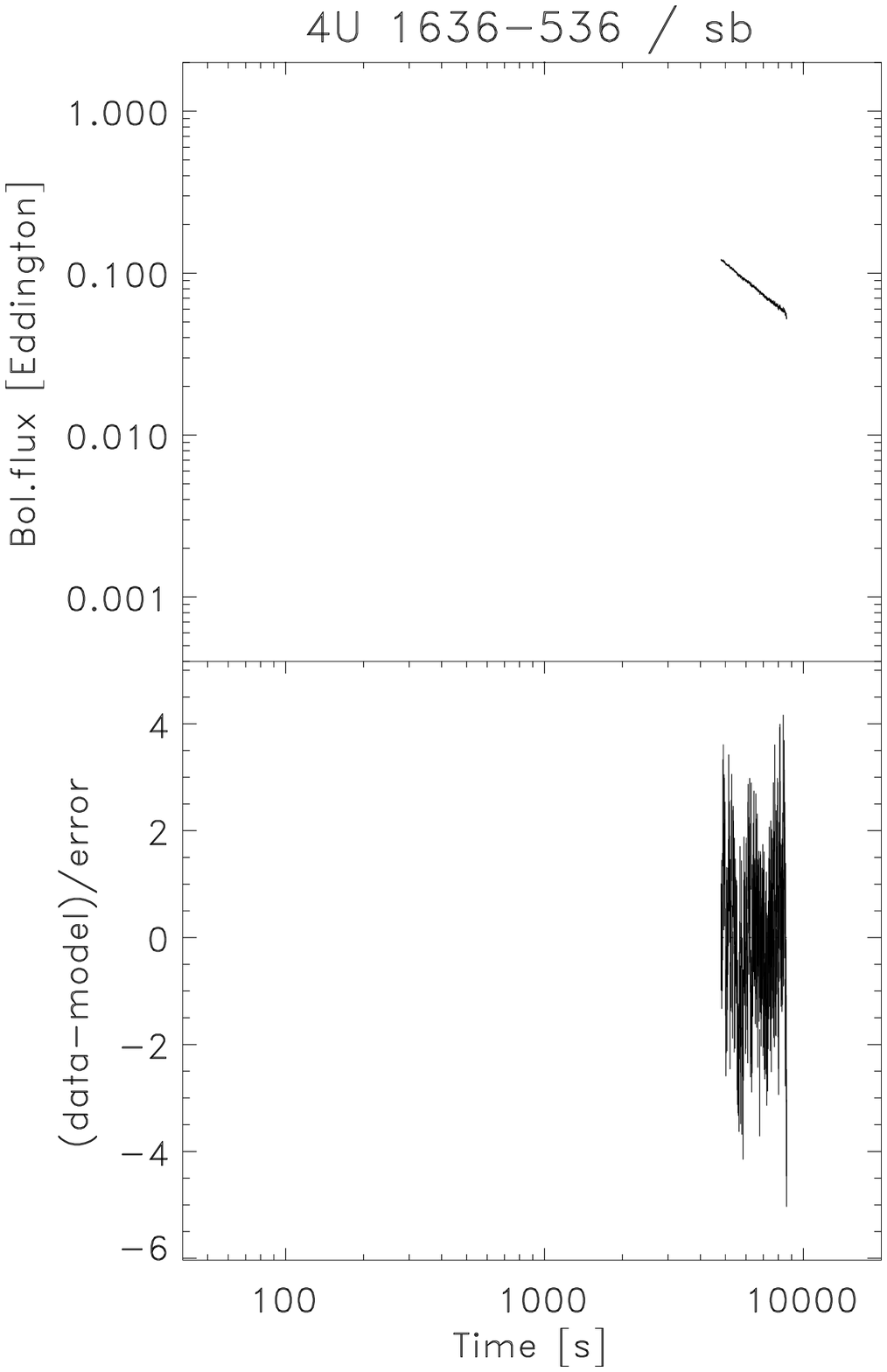}
\caption{Power-law fits to bolometric flux histories of 16
  representative bursts from 14 different LMXBs. The X-axis represents
  time since burst onset. There are two panels per burst. The upper
  panels show the bolometric flux in units of the Eddington flux, as
  determined from the highest peak flux over all bursts observed per
  source. The lower panels show the deviation with respect to the
  model in units of $\sigma$ per data point. All axes of the upper
  panels and time have an identical dynamic ranges so that slopes can
  be directly compared. The burst are identified in a slightly
  abbreviated but straightforward manner (c.f., Table \ref{tab1}). The
  last two bursts were added to extend the time range, although they
  do not have smooth decays.
\label{fig:fits}}
\end{figure*}

Figure~\ref{fig:fits} shows for a representative subset of all bursts
the power-law fits to the bolometric flux data.

\begin{figure}[t]
\includegraphics[width=0.88\columnwidth,angle=0]{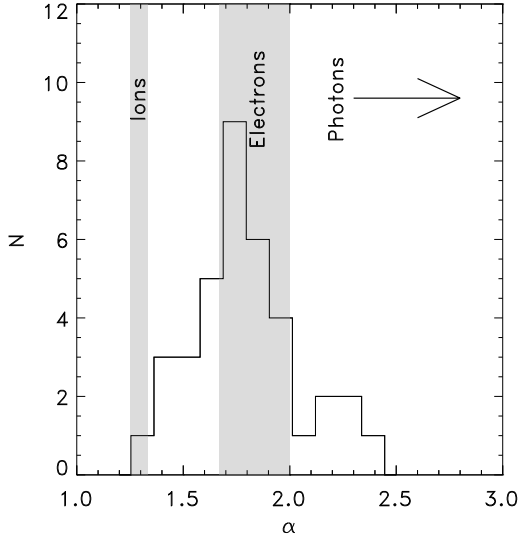}
\caption{Histogram of power-law decay index of fits to to bolometric
  flux history of all 37 bursts. The gray areas indicate the canonical
  values for a pure ideal ion gas (1.25-1.33; see text) and a pure
  degenerate electron gas (1.67-2.00). When the photons contribute
  significantly to the heat capacity, $\alpha$ can become higher.
\label{fig:his}}
\end{figure}

We verified the robustness of the power-law fits to the photon count
rate data in order to obtain a sense of the possible systematic errors
of the power-law decay index. This verification encompassed 3 tests:
how does the power law change when leaving free $F_{\rm bg}$, when
limiting the data to fluxes that are 2\% or higher than the peak flux,
or when both these tests were applied at the same time. We find that
individual values of the decay index change on average by 3 to 4\% and
that the mean value over all bursts changes by only 0.2\%. Therefore,
the result seems robust.

Concentrating on the power-law fits to the bolometric flux histories
of the 35 ordinary bursts, we find that the power-law decay index lies
between 1.4 and 2.4, with a weight average of $1.844$ and a standard
deviation of 0.24. This is a 0.511 steeper index than the derived
value in \S\ref{sec:intro} (4/3).  The histogram (Fig.~\ref{fig:his})
looks bimodal with a primary distribution between $\alpha=1.4$ and 2.1
peaking at $\alpha\approx1.75$ and a tail of 5 bursts with
$\alpha>2.1$.  The weight averages of the 4 objects with multiple
bursts are: $1.71\pm0.11$ for 4U 1608-52, $1.87\pm0.16$ for SAX
J1808.4-3658, $2.18\pm0.26$ for HETE J1900.1-2455 and $1.64\pm0.16$
for Aql X-1. Due to the large uncertainties (these are the standard
deviations), there is no strong evidence for systematic differences
between objects. The decay index for the two long bursts is low. That
of the superburst is the only one consistent with the canonical 4/3
value.

If we exclude the 13 bursts that we took under reserve in
Sect.~\ref{sec:modeling}, then the average over the 22 ordinary bursts
is $\alpha=1.75\pm0.21$. This average is only marginally shallower and
still 0.418 steeper than 4/3. The range of $\alpha$ also remains
similar: 1.4--2.3. Comparing H-rich against H-poor accretors we find
1.86$\pm$0.26 against 1.81$\pm$0.15, which is an insignificant
difference.

Many of the power-law fits to the photon count rate data are of good
quality as well, with a weighted average of the power-law decay index
of $1.929$, which is only 0.085 different from the value for the
bolometric flux, and a standard deviation of 0.16 - smaller than for
the bolometric flux. On a burst-by-burst basis, the difference is
larger than 0.1 seven times, most notably in 4U0512-40/2, 4U
1705-44/77 (these are the two ordinary bursts with the shallowest
power-law decay index in bolometric flux), 4U 1608-52/9 and Aql
X-1/11.

\section{Discussion}
\label{sec:discussion}

For 35 ordinary 'clean' PCA-detected X-ray bursts, that have the
simplest light curve shape (complete coverage, monotonic and smooth
decay, non-variable non-burst emission and high peak flux to pre-burst
flux ratio), we find that a power law is always a better description
of the decay portion than an exponential function, whether that decay
is measured in units of photon count rate or bolometric flux. The same
applies for 2 additional long, but not so 'clean' bursts (i.e., they
show smooth decays for only part of the decay).  This preference for
the power law confirms the theoretical expectations for the cooling
curve \citep[e.g.,][]{cum04} and warns against the common use of a
single exponential function. The power-law decay index is considerably
steeper than the canonical 4/3 value (c.f., Eq.~\ref{eqn2}) for the 35
ordinary bursts and variable from burst to burst.  What could be the
reason for this fast cooling and spread in ordinary bursts?  We first
consider whether systematic effects in the data analysis could bias
the measured power-law indices, and then show that a more careful
consideration of the microphysics of the heat capacity of the cooling
layer naturally gives a steeper decay than the 4/3 law predicted by
assuming a constant heat capacity and constant opacity.

\subsection{Systematic effects}

As mentioned above, $\alpha$ is strongly correlated with $t_s$, so if
$t_s$ were wrong, that would introduce a systematic offset in
$\alpha$. An offset of 1 s, for our bursts, translates to a change in
$\alpha$ of 0.05 (for an illustration of the coupling between both
parameters, see Fig.~\ref{fig:chi}). However, in order to get
shallower index values, one would need to introduce values for $t_s$
that are later in the burst, in other words cooling would need to
start later than the end of the nuclear burning. That seems an
unlikely scenario. Furthermore, the power-law fits become unacceptable
(see Fig.~\ref{fig:chi}).

In order to estimate the bolometric flux, the empirical Planck
function is assumed to apply outside the 3 to 20 keV bandpass. If that
assumption is increasingly wrong with decreasing temperatures, that
would introduce a bias and change in power-law decay index. The lowest
measured temperature is 0.8 keV. The peak of the energy spectrum is
then just below the lower threshold of the bandpass and the risk for
wrong extrapolations the largest. There is a rich body of literature
about the deviation of NS photosphere models from the Planck spectrum.
These all agree that the ratio between color temperature, which is the
fitted black body value, and the effective temperature, which would be
a fair representation of the Planck spectrum, is greater than 1 and
decreases with color temperature
\citep[e.g.,][]{lon86,mad04,sul12}. We tested the effect on our
analysis by employing the model of \cite{sul12}, calculating the
'true' bolometric flux according to the model, simulating the spectrum
for a range of temperatures, fitting a black body model between 3 and
20 keV with the RXTE response matrix and calculating the bolometric
flux from that. We find that the true bolometric flux is always larger
than the one derived from the black body fit, that this deviation
increases towards lower temperatures, but that it remains limited to
10\% at 0.8 keV (1.6\% at 2.1 keV). This difference is by far (by
about factor of 10) insufficient to explain the difference in
power-law decay index. The non-Planckian nature of the burst spectrum
alone cannot explain the discrepancy between the measured and
predicted power-law decay index.

\subsection{Intrinsic effects}

The 4/3 decay index (Eq.~\ref{eqn2}) is derived from the assumptions
that $C_{\rm P}$ is constant and independent of the temperature in the
layer $T$ (Eq.~\ref{eqn01}) and that the cooling luminosity of the
layer is $\propto T^4$ (Eq.~\ref{eqn02}). In fact, a more detailed
consideration of the microphysics shows that both of these assumptions
must be modified for the neutron star outer layers.

First, we consider the heat capacity $C_{\rm P}$. The heat capacity is
independent of $T$ for an ideal gas, but we know that the ignited
layer of plasma consists of two components: the ions, which can be
considered an ideal gas, and the electrons, which are degenerate
beyond a certain depth. Comparing the thermal energy $k_BT$ to the
non-relativistic Fermi energy $E_F=(\hbar^2/2m_e)(3\pi^2n_e)^{2/3}$,
the electrons are degenerate ($k_B T <E_F$) for densities greater than
$1.2\times 10^4\ {\rm g\ cm^{-3}} T_8^{3/2} (Y_e/0.5)^{-1}$ or column
depths greater than about 10$^6$~g~cm$^{-2} T_8^{3/2} (Y_e/0.5)^{-1}$,
where $Y_e$ is the electron number fraction and $T_8=T/10^8\ {\rm
  K}$. Ignition depths for X-ray bursts are typically a factor of
10$^2$ deeper in column depth \citep[e.g.,][]{cum00}.  The heat
capacity of degenerate electrons is
\begin{eqnarray}
C_{P,e} & = & {\pi^2\over 2}Y_e {k_B^2T\over m_p E_F}\propto T
\label{eqn3}
\end{eqnarray}
For densities greater than $\rho\approx 10^7\ {\rm g\ cm^{-3}}$
(column depths $\gtrsim 10^{10}\ {\rm g\ cm^{-2}}$), the electrons are
relativistically degenerate, in which case the prefactor in the heat
capacity is $\pi^2$ rather than $\pi^2/2$, but the scaling is still
$C_{P,e}\propto T$.

The total heat capacity is the sum of the contributions from ions,
electrons, and radiation. At low temperature, the ions dominate the
heat capacity giving $C_{\rm P}$ approximately constant. At higher
temperatures, the electron heat capacity increases and eventually
dominates, so that the total heat capacity becomes proportional to
temperature. A specific heat capacity that is proportional to $T$
changes Eq.~\ref{eqn1} to $T\propto t^{-1/2}$ and Eq.~\ref{eqn2} to
$L\propto t^{-2}$. In general, if $C_{\rm P}\propto T^\beta$, $\alpha
= \frac{4}{3-\beta}$. Thus, for a mixture of ideal and degenerate gas
$\alpha$ is expected to range between 4/3 and 2. If the temperature is
relatively low, it will remain near 4/3, but if it is high the heat
capacity of the electrons increases while that of the ionic gas
remains constant and $\alpha$ will grow. At higher temperatures still,
radiation pressure becomes non-negligible with respect to the gas
pressure, and $\alpha$ will increase even further because the heat
capacity of a pure photon gas is $\propto T^3$ at constant volume (and
formally divergent at constant pressure) and $\alpha$ grows to
infinity \citep[e.g.,][]{cum00}.

\begin{figure}[t]
\includegraphics[width=1.0\columnwidth]{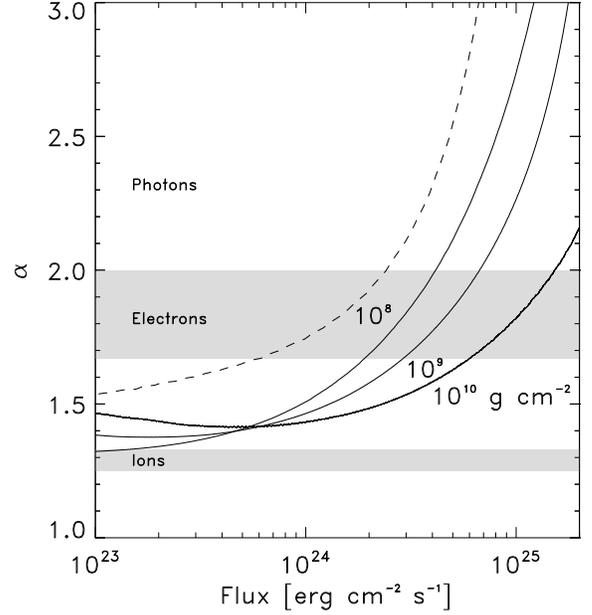}
\caption{The light curve slope as a function of flux as calculated in
  multizone models of a cooling layer (solid curves) with column
  depths $y=10^8$, $10^9$, and $10^{10}\ {\rm g\ cm^{-2}}$. For
  comparison we show the one-zone model result for $y=10^8\ {\rm
    g\ cm^{-2}}$ as the dashed curve. The gray areas show the expected
  range of values of $\alpha$ when ions dominate the heat capacity or
  when electrons dominate the heat capacity, taking the cooling to lie
  between $L\propto T^4$ and $L\propto T^5$.
\label{fig:andrew}}
\end{figure}

The dependence of the cooling luminosity on the layer temperature
depends on the details of the temperature profile in the layer,
connecting the temperature near the base of the layer to the
temperature at the photosphere. For a constant flux, this relation is
determined by integrating the radiative diffusion equation
\begin{eqnarray}
F &=& -{4a c T^3\over 3\kappa\rho}{dT\over dr}
\label{eqn4}
\end{eqnarray}
where $\kappa$ is the opacity. For constant opacity, the integration
gives $F\propto T^4$, but the scaling with temperature can be
different when the opacity is temperature and depth dependent. For
example, the relation between surface temperature $T_{\rm eff}$ and
the temperature deep in the crust at densities of $\rho \gtrsim
10^{10}\ {\rm g\ cm^{-3}}$ is $T_{\rm eff}\propto T^{2.2}$
\citep{gud82}. We calculated a series of constant flux temperature
profiles in the neutron star envelope to determine the scaling of
luminosity with temperature at the base of the layer, $L\propto
T^\gamma$. We find that for column depths typical of X-ray burst
ignition, $10^8$--$10^9\ {\rm g\ cm^{-2}}$, $\gamma\approx 4$--$5$.

Following through the argument leading to Eq.~\ref{eqn2} for $L\propto
T^\gamma$ gives $\alpha = \gamma/(\gamma-1-\beta)$. Therefore for the
range $\gamma=4$--$5$, we expect $\alpha=1.25$--$1.33$ when ions
dominate the heat capacity ($\beta=0$), $\alpha=1.67$--$2.0$ when
electrons dominate the heat capacity, and larger values of $\alpha$
when radiation makes a significant contribution. These values are
shown as gray regions in the histogram of $\alpha$ in
Figure~\ref{fig:his}. We see that there is a good match to the
observed values of $\alpha$ when degenerate electrons or radiation is
taken into account.

To further investigate the agreement between the predicted and
observed values of $\alpha$, we calculated the expected values of
$\alpha$ as a function of the flux from the star. We did this in two
ways. The first is a one-zone model based on the argument leading to
Eq.~\ref{eqn2}. For a given flux $F$ and layer column depth $y$, we
first use our constant flux envelope models to find the temperature at
the base of the layer. We calculate the heat capacity temperature
scaling $\beta$ using the base temperature and base pressure $P=gy$,
where $g$ is the gravitational acceleration (where we use the fact
that the layer is in hydrostatic balance). To validate the one-zone
approach, we also calculated a series of multizone cooling models
following \cite{cum04} but extended to shallower layers. In these
models, we heated a layer of a given depth by depositing the amount of
energy expected from complete helium burning ($1.6\ {\rm MeV}$ per
nucleon) and then followed the cooling of the layer by integrating the
thermal diffusion equation. We then calculated the local slope of the
light curve $\alpha=-d\ln L/d\ln t$ as a function of time and
therefore as a function of flux.

The results of the multizone models are shown as solid curves in
Fig.~\ref{fig:andrew} for column depths $y=10^8$, $10^9$, and
$10^{10}\ {\rm g\ cm^{-2}}$. The one-zone model results for
$y=10^8\ {\rm g\ cm^{-2}}$ are shown as the dashed curve. The shape of
the one-zone model curve matches the multizone model well, but the
one-zone model decay is everywhere steeper than found in the multizone
model. The reason for this is that in the multizone model, which
follows the temperature profile of the envelope in detail, a
significant amount of heat is conducted inwards as the layer cools, so
the effective mass of the cooling layer changes. This is not taken
into account in the one-zone model which assumes a fixed column depth
$y$ and therefore cools faster (larger $\alpha$).

Fig.~\ref{fig:andrew} shows that the expected behavior is that
$\alpha$ will decrease with flux (light curve decay becomes
shallower). The reason for this is that initially when the layer is
hot, radiation pressure is significant, but later the heat capacity
becomes dominated by the electrons and ions.  At larger column depths,
the influence of both radiation pressure and electrons is smaller, and
a smaller range of values of $\alpha$ is explored. The decreasing
influence of degenerate electrons towards larger ignition depths comes
about because only a fraction $k_{\rm B}T/E_{\rm F}$ of the degenerate
electrons near the Fermi surface participates in the thermal energy,
and $E_F$ increases as the layer becomes thicker.

Contrary to the models, we do not detect a change in $\alpha$ in the
data, because all data per burst are consistent with a single power
law. For bursts for which the fitted data do not cover the upper
decade in flux, this is not unexpected. Most of the change in $\alpha$
is in that range. However, around half the bursts do cover that upper
range. Thus, the model appears insufficient. It may be that the base
temperature is ill constrained due to insufficient knowledge about the
total energy liberated. For example, if the energy per nucleon is 0.6
MeV instead of 1.6 MeV as assumed, which is the number for helium
burning to carbon instead of iron, $\alpha$ will remain below 2.05 for
$y=10^8$~g~cm$^{-2}$ for a flux below
10$^{25}$~erg~cn$^{-2}$s$^{-1}$. This compares to $\alpha<2.8$ for an
energy release of 1.6~MeV per nucleon. Measuring this is difficult for
PRE bursts, because a significant fraction of the energy is
invisible. This may be a subject of future refined modeling.

\begin{figure}[t]
\includegraphics[width=0.9\columnwidth,angle=0]{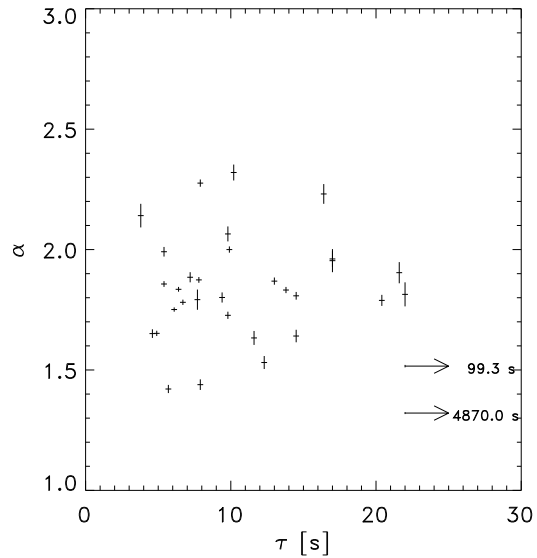}
\caption{Measurements of $\alpha$ versus $\tau$. Horizontal bars are
  simple markers (exponential function are bad fits to the data) and
  vertical bars $1\sigma$ errors of $\alpha$. There is no clear
  correlation between both parameters, except that the longest burst
  has the lowest value.
\label{fig:alphavsduration}}
\end{figure}

The spread of $\alpha$ from burst to burst is in line with the above
explanation with degenerate electrons and photons, because one expects
a spread in ignition conditions from burst to burst. In principle,
$\alpha$ may constrain the ignition conditions. For instance, a high
$\alpha$ points to a low ignition depth. We tried to verify a
dependence between $\alpha$ and ignition depth, by assuming that burst
duration as measured with $\tau$ depends monotonically on ignition
depth \citep[see][]{cum04}. Figure~\ref{fig:alphavsduration} shows
$\alpha$ versus $\tau$.  There is no correlation between both
parameters, except that the longest (super) burst has the lowest
$\alpha$, which is consistent with $\alpha=4/3$. The absence of
correlation for ordinary and intermediate duration bursts is probably
due to the fact that $\tau$ is not an accurate enough proxy for
ignition depth. More detailed light curve modeling that includes early
times in intermediate duration bursts is necessary to make a more
accurate verification.

\section{Conclusion}

We have verified, for a representative set of 35 ordinary
thermonuclear X-ray bursts from 14 neutron stars, that the radiative
decay follows a power law rather than an exponential decay function,
and find that the decay index of the power law is steeper than seen in
long superbursts (1.8 versus 1.3). Also, it varies from burst to
burst, even if from the same neutron star. We hypothesize that this is
due to the influence of degenerate electrons and photons on the heat
capacity of the ignited layer.

We are unable to confirm this hypothesis through independent
measurements of ignition depths or through detection of a change in
$\alpha$. That will only be possible through more complete modeling of
burst light curves, particularly at early phases when the cooling wave
is traveling from the photosphere to the ignition depth. That is not
straightforward, because data from that phase suffer from the effects
of photospheric expansion. Therefore, the model will have to include
those effects. Currently, there are no such models. If it would become
possible to confirm the relationship between ignition depth and
$\alpha$ for a number of bursts, measurement of $\alpha$ might yield a
valuable constraint on ignition depth.

\acknowledgements We thank Laurens Keek, Duncan Galloway and Nevin
Weinberg for useful discussions and Erik Kuulkers for providing the
data of the superburst of 4U 1636-536. AC is supported by an NSERC
Discovery Grant and is an associate member of the CIFAR Cosmology and
Gravity Program.

\bibliographystyle{aa} \bibliography{references}

\end{document}